\documentclass[aps,pre,twocolumn,superscriptaddress,showpacs]{revtex4-1}

\usepackage{epsfig}
\usepackage{epstopdf}
\usepackage{amsmath}
\usepackage{amssymb}

\begin{document}

\newcommand{\bra}{\left\langle}
\newcommand{\ket}{\right\rangle}

\title{Normal mode analysis of spectra of random networks}

\author{G. Torres-Vargas}
\email[]{gamaliel\_torres@uaeh.edu.mx}
\affiliation{Instituto de Ciencias B\'asicas e Ingenier\'ia, Universidad Aut\'onoma del Estado de Hidalgo, Pachuca 42184, Hidalgo, Mexico}
\affiliation{Posgrado en Ciencias Naturales e Ingenier\'ia, Universidad Aut\'onoma Metropolitana Cuajimalpa, 05348 CDMX, Mexico}

\author{R. Fossion}
\email[]{fossion@nucleares.unam.mx}
\affiliation{Instituto de Ciencias Nucleares, Universidad Nacional Aut\'onoma de M\'exico, 04510 CDMX, Mexico}
\affiliation{Centro de Ciencias de la Complejidad (C3), Universidad Nacional Aut\'onoma de M\'exico, 04510 CDMX, Mexico}

\author{J. A. M\'endez-Berm\'udez}
\email[]{jmendezb@ifuap.buap.mx}
\affiliation{Instituto de F\'isica, Benem\'erita Universidad Aut\'onoma de Puebla, Apartado Postal J-18, Puebla 72570, Mexico}

\begin{abstract}
Several spectral fluctuation measures of random matrix theory (RMT) have been applied in the study of spectral properties of networks. However, the calculation of those statistics requires performing an unfolding procedure, which may not be an easy task. In this work, network spectra are interpreted as time series, and we show how their short and long-range correlations can be characterized without implementing any previous unfolding. In particular, we consider three different representations of Erd\H{o}s-R\'enyi (ER) random networks: standard ER networks, ER networks with random-weighted self-edges, and fully random-weighted ER networks. In each case, we apply singular value decomposition (SVD) such that the spectra are decomposed in trend and fluctuation normal modes. We obtain that the fluctuation modes exhibit a clear crossover between the Poisson and the Gaussian orthogonal ensemble statistics when increasing the average degree of ER networks. Moreover, by using the trend modes, we perform a data-adaptive unfolding to calculate, for comparison purposes, traditional fluctuation measures such as the nearest neighbor spacing distribution, number variance $\Sigma^2$, as well as $\Delta_3$ and $\delta_n$ statistics. The thorough comparison of RMT short and long-range correlation measures make us identify the SVD method as a robust tool for characterizing random network spectra.
\end{abstract}

\maketitle

\section{Introduction}
\label{SecI}

Recently, spectra of ordered eigenvalues of finite random matrices have been studied as time series \cite{Fossion13,Torres17,Torres18}. In particular, by applying the singular value decomposition (SVD) to standard Gaussian ensembles~\cite{Fossion13,Torres17}, and also to nonstandard random-matrix ensembles~\cite{Torres18}, a decomposition of each spectrum, in trend and fluctuation normal modes was realized. In this way, a characterization of the transition from Gaussian orthogonal ensemble (GOE) to Poisson statistics, was obtained in a direct way without performing the technical step known as \emph{unfolding}. The above was achieved based on the behavior of the fluctuation modes, which can be scale invariant and follow a power law, or exhibit a crossover between those two limits.

In the framework of random-matrix theory (RMT), the unfolding procedure has the purpose of separating the smooth global energy level density, $\overline{\rho}(E)$, from the local fluctuations, $\widetilde{\rho}(E)=\rho(E)-\overline{\rho}(E)$, and rescaling the level spacing to unit mean spacing. After unfolding, the correlations in the spectra can be quantified by calculating spectral fluctuation measures such as the nearest-neighbor spacing distribution (NNSD), for short-range correlations, and the number variance $\Sigma^2$, as well as $\Delta_3$ and $\delta_n$ statistics, for long-range correlations~\cite{meh91,rel02,Riser17}. However, it is known that the statistical results obtained can be quite sensitive to the specific implemented unfolding procedure~\cite{gom02,abu12,abu14,ber16,she18}. In this manner, applying SVD directly to the spectra is an unambiguous technique for studying spectral correlations without performing any unfolding procedure, avoiding the introduction of possible artifacts~\cite{Torres17,Torres18}. Besides, in order to calculate the standard spectral fluctuation measures from RMT, a \emph{data-adaptive} unfolding can be realized. This unfolding method has the advantage of being a self-consisted method, defined in an intrinsic way, such that the global part of the spectra is obtained from the data itself, instead of being imposed extrinsically~\cite{Fossion13}.


In this context, one of the fields where RMT has found new applications is in the study of eigenspectra of adjacency matrices of networks, see e.g. Refs.~\cite{ZX00,ZX01,GKK01,F01,F02,DGMS03,GGS05,SKHB05,AM05,PV06,GT06,JB07,BJ07,JKBH08,JB08,ZYYL08,J09,JB09,CJH09,GGS09,JSVL10,JZL11,S12,YWG15,net01,net02,DGK17,SJ18,net03,net04,net05}. Since random networks models can be represented by sparse random-matrix ensembles, a natural application of SVD would be to the study of the spectral properties of such models. In particular, in the present work, we focus on Erd\H{o}s-R\'enyi (ER) networks. The ER random graph model was introduced by Solomonoff and Rapoport~\cite{SR51}, and later studied into detail by Erd\H{o}s and R\'enyi~\cite{ER59,ER60}. This model is also known as uncorrelated random graph model. In the standard ER model a random graph $G(N,\alpha)$ consists of $N$ vertices and each edge appears independently with probability $\alpha\in[0,1]$. Indeed, it is known that the ER model depends on the single parameter $\xi$~\cite{jac01,AB02,net01}, i.e., $G(N,\alpha)\equiv G(\xi)$, where
\begin{equation}
\xi=N\alpha
\label{xi}
\end{equation}
is the graph average degree. Here, to apply SVD on rather well known random networks, we consider three representations of the ER model: (a) fully random-weighted ER networks, (b) ER networks with random-weighted self-edges, and (c) standard ER networks. We define these variations of ER networks in Sec.~\ref{SecIV}.

The paper is organized as follows. In Sec.~\ref{SecII}, we describe the SVD method and show how it decomposes the network spectra in trend and fluctuation modes. Furthermore, we present a short-range spectral fluctuation measure, the distribution $P(\tilde{r})$, which also does not require any previous unfolding. In Sec.~\ref{SecIII}, we expose briefly the standard spectral fluctuation measures employed in RMT. In Sec.~\ref{SecIV}, we present the results of applying SVD to the aforementioned representations of ER networks, and also obtain the $P(\tilde{r})$ distribution in each case. Finally, we perform a data-adaptive unfolding of the spectra in order to calculate the NNSD, $\Sigma^2$, $\Delta_3$ and $\delta_n$ statistics, and compare these results with those obtained previously. In Sec.~\ref{Conclusions}, we give our conclusions.

\section{Spectral fluctuation analysis without unfolding}
\label{SecII}

\subsection{Singular Value Decomposition}

Let us consider $m$ eigenspectra (time series), each with $n$ ordered eigenvalues, where $m=1,\ldots,M$ and $n=1,\ldots,N$. We can construct a $M\times N$-dimensional matrix (multivariate time series), $\mathbf{X}$, if we arrange the eigenspectra, $X^{(m)}(n)$, in the following way:
\begin{equation}
\mathbf{X}
=\left(\begin{array}{cccc}
  X^{(1)}(1) & X^{(1)}(2) & \ldots & X^{(1)}(N) \\
  X^{(2)}(1) & X^{(2)}(2) & \ldots & X^{(2)}(N) \\
  \vdots & \vdots & \ddots & \vdots \\
  X^{(M)}(1) & X^{(M)}(2) & \ldots & X^{(M)}(N) \\
\end{array}\right).
\label{A1}
\end{equation}
SVD is a matrix decomposition technique that expresses $\mathbf{X}$, in a unique and exact way, as the product of three matrices
\begin{equation}
\mathbf{X}=\mathbf{U}\boldsymbol{\Sigma}\mathbf{V}^T,
\label{A2}
\end{equation}
where $\mathbf{U}$ and $\mathbf{V}$ are square matrices of dimension $M \times M$ and $N \times N$, respectively. $\boldsymbol{\Sigma}$ is a $M \times N$-dimensional matrix with diagonal elements only, which are the ordered \emph{singular values} $\sigma_1\geq\sigma_2\geq\ldots\geq\sigma_r$, with $r= \textmd{rank}(\mathbf{X})\leq \min[M,N]$. Here, the superscript $T$ indicates matrix transpose. Equivalently, the decomposition of Eq.~(\ref{A2}) can be expressed as a weighted sum of elementary matrices (rank one matrices), $\mathbf{X}_k$, in the following way
\begin{equation}
\mathbf{X} = \sum_{k=1}^r \sigma_k \mathbf{X}_k,
\label{A3}
\end{equation}
where $\mathbf{X}_k=\vec{u}_k\vec{v}_k^T$, and $\vec{u}_k$ are the orthonormal vectors constituting the $k$th column of the matrix $U$, while $\vec{v}_k$ are the corresponding orthonormal vectors constituting the $k$th column of the matrix $V$. Here, $\vec{u}_k\vec{v}_k^T\equiv\vec{u}_k\otimes\vec{v}_k$ denotes the outer product of $\vec{u}_k$ and $\vec{v}_k$. The vectors $\vec{u}_k$ are called \emph{left-singular vectors} of $\mathbf{X}$, which span its column space, and can be interpreted as \emph{projection coefficients}. On the other hand, the vectors $\vec{v}_k$ are called \emph{right-singular vectors} of $\mathbf{X}$, which span its row space, and constitute an orthonormal basis for the time series of the ensemble $\mathbf{X}$, which is why they are known as \emph{normal modes}. The set $\{\sigma_k,\vec{u}_k,\vec{v}_k\}$ is called \emph{eigentriplet}, and it defines completely the eigenmode of order $k$. Therefore, any eigenspectrum, $X^{m}(n)$, can be expressed as a superposition of the normal modes, in the following way
\begin{equation}
X^{(m)}(n)=\sum_{k=1}^r\sigma_kU_{mk}\vec{v}_k^T(n),
\label{A4}
\end{equation}
where the matrix elements $U_{mk}$ serve as coefficients in this projection, and $\sigma_k$ can be interpreted as weights that distinguish between \emph{trend} and \emph{fluctuation} modes. The square of a singular value, $\lambda_k={\sigma_k}^2$, is the \emph{partial variance}, which gives the fraction of the total variance, $\lambda_{tot}=\sum_{k=1}^r\lambda_k$, of the multivariate time series $\mathbf{X}$ carried by the normal mode $\vec{v}_k$. In this context, the trend modes are usually characterized by very large partial variances, while the fluctuation modes are associated with much smaller partial variances. Therefore, the number of normal modes to be included in the trend, $n_T$, can be easily identified from a log-log plot of the ordered partial variances, known as \emph{scree diagram}. The global properties (trend) of a spectrum are represented by \emph{nonoscillating} normal modes, $\vec{v}_k$, associated to the first few dominant partial variances $\lambda_k$, $k=1,\ldots,n_T$. On the other hand, the local properties (fluctuations) corresponds to \emph{oscillating} normal modes associated to the rest of partial variances $\lambda_k$, with $k=n_T+1,\ldots,r$~\cite{Fossion13}. Thus, we can separate the eigenspectrum $X^{(m)}(n)$ in a global and local part, $\overline{X}^{(m)}(n)$ and $\widetilde{X}^{(m)}(n)$, respectively,
\begin{equation}
X^{(m)}(n)=\overline{X}^{(m)}(n)+\widetilde{X}^{(m)}(n),
\label{A5}
\end{equation}
where, in terms of Eq.~(\ref{A4}), they are given by
\begin{align}
\overline{X}^{(m)}(n)&=\sum_{k=1}^{n_T}\sigma_kU_{mk}\vec{v}_k^T(n),\label{A6}\\
\widetilde{X}^{(m)}(n)&=\sum_{k=n_T+1}^r\sigma_kU_{mk}\vec{v}_k^T(n).\label{A7}
\end{align}
In Sec.~\ref{SecIII} the trend obtained from Eq.~(\ref{A6}) will be employed to perform a data-adaptive unfolding of the eigenspectra of ER networks. A more detailed explanation of SVD, and the data-adaptive unfolding, can be found in~\cite{tor14,fos14,fos15}. It has been shown that for standard Gaussian ensembles of RMT, the part of the scree diagram corresponding to the higher-order normal modes (fluctuation modes) follows a power law~\cite{Fossion13,Torres17},
\begin{equation}
\lambda_k\propto1/k^\gamma,
\label{B1}
\end{equation}
where $k=n_T+1,\ldots,r$. Given the scale invariance of this power law, the values of the exponent $\gamma$ do not depend on the size of the ensemble, being $\gamma=1$ in the GOE limit and $\gamma=2$ in the Poisson limit. In this way, the exponent $\gamma$ serves to characterize the long-range correlations present in the fluctuations of the eigenspectra. When nonstandard random-matrix ensembles are considered, such as the $\beta$-Hermite ensemble and the sparse matrix ensemble, the scale invariance of the fluctuations is lost. In these cases, a crossover, instead of a power law, is observed in the scree diagrams~\cite{Torres18}.

\subsection{Distribution P$(\tilde{r})$}

A short-range spectral fluctuation measure which, like the scree diagram, does not require to perform a previous unfolding procedure, is the distribution $P(\tilde{r})$~\cite{oga07,ata13}, where
\begin{equation}
\tilde{r}_n=\frac{\min(s_n,s_{n-1})}{\max(s_n,s_{n-1})}
\label{B3}
\end{equation}
is the ratio of two consecutive eigenvalue spacings,
\begin{equation}
s_n=X(n+1)-X(n).
\label{B4}
\end{equation}
In this way, the $P(\tilde{r})$ distribution does not depend on the local density of states. We will compare the results obtained for $P(\tilde{r})$, with those obtained for an equivalent and widely used short-range spectral fluctuation measure of RMT, the NNSD. Besides, we will calculate the average value of $\tilde{r}_n$, $\langle\tilde{r}\rangle$, in order to characterize the transition from Poisson to GOE statistics, as $\xi$ increases. In particular, in the Poisson and GOE limits the $P(\tilde{r})$ distributions are given, respectively, by~\cite{ata13}
\begin{equation}
P(\tilde{r})=\frac{2}{(1+\tilde{r})^2} \, \Theta(1-\tilde{r}),
\label{Pofr_P}
\end{equation}
and
\begin{equation}
P(\tilde{r})=\frac{27}{4}\frac{\tilde{r}+\tilde{r}^2}{(1+\tilde{r}+\tilde{r}^2)^{5/2}} \, \Theta(1-\tilde{r}).
\label{Pofr_GOE}
\end{equation}

\section{Spectral fluctuation measures of random matrix theory}
\label{SecIII}

In this section, we describe the traditional \emph{spectral fluctuation measures} used in RMT. Such measures describe the correlations between two or more \emph{unfolded} eigenvalues (energy levels) of a random-matrix spectrum. Some of the statistics most commonly used are the NNSD, the number variance $\Sigma^2$, as well as the $\Delta_3$ and $\delta_n$ statistics~\cite{meh91,rel02,Riser17}. The NNSD describes the correlations between two consecutive levels; this is why the NNSD is said to be a \emph{short-range} spectral fluctuation measure. Other measures such as $\Sigma^2$ and $\Delta_3$ describe the correlation properties among a larger number of energy levels; thus, they are known as \emph{long-range} spectral fluctuation measures. In order to calculate such measures, it is necessary to perform an unfolding of the eigenspectra (energy spectra), $X^{(m)}(n)$. With this purpose we perform a data-adaptive unfolding, employing the corresponding global part of the ER network spectra, obtained when we applied SVD, Eq.~(\ref{A6}). A detailed explanation of this data-adaptive unfolding can be found in Refs.~\cite{tor14,fos14}. After performing the unfolding, spectra $x(n)$ with a mean spacing equal to one, $\langle s(n)\rangle=1$, are obtained, where
\begin{equation}
s(n)=x(n+1)-x(n).
\label{C1}
\end{equation}
Note that $s(n)\neq s_n$ of the previous section, Eq.~(\ref{B4}), and unlike $X(n)$, the unfolded spectra, $x(n)$, have a uniform distribution.

\subsection{$\delta_n$ statistics}

A long-range spectral fluctuation measure equivalent to the scree diagram is the $\delta_n$ statistics~\cite{rel02,Riser17}. It measures the deviation of the unfolded eigenvalue from its mean value $n$, and it is defined by
\begin{equation}
\delta_n=\sum_{i=1}^{n}(s(i)-\langle s\rangle) ,
\label{C2}
\end{equation}
where $n=1,\ldots,N-1$. In Ref.~\cite{rel02} the function $\delta_n$ was considered as a discrete and finite time series. It was found that for energy spectra of regular and chaotic quantum systems, the corresponding power spectrum, $P(f)=|\boldsymbol{\hat{\delta}}_f|^2$, where $\boldsymbol{\hat{\delta}}_f$ is the Fourier transform of $\delta_n$,
\begin{equation}
\boldsymbol{\hat{\delta}}_f=\frac{1}{\sqrt{N}}\sum_n\delta_n\exp\left(-2\pi i{\frac{nf}{N}}\right),
\label{C3}
\end{equation}
follows a power law,
\begin{equation}
P(f)\propto1/f^{\beta},
\label{C4}
\end{equation}
which indicates that the spectral fluctuations are scale invariant (fractal). For Poisson spectra $\beta=2$, while $\beta=1$ for the GOE ensemble, and the rest of classical Gaussian ensembles; the Gaussian unitary ensemble (GUE), and the Gaussian symplectic ensemble (GSE).

\subsection{Nearest neighbor spacing distribution (NNSD)}

The NNSD shows how the spacings, $s(n)$, between consecutive unfolded energy levels fluctuate around the average spacing, $\langle s\rangle=1$. The distribution $P(s)$ is defined as the probability density that two adjacent unfolded levels are separated by a distance $s$. After we have unfolded an energy spectrum, the NNSD is obtained by calculating the spacings between consecutive levels, Eq.~(\ref{C1}). The distribution exhibited by the spacings, $s$, will be determined by its correlation properties. For example, if the positions of the energy levels are not correlated, the $P(s)$ follows a Poisson distribution,
\begin{equation}
P(s)=\exp(-s).
\label{C5}
\end{equation}
Note that the Poisson distribution has its maximum at $s=0$, indicating that small spacings have a higher probability of occurrence (there is no repulsion of levels). On the other hand, the unfolded levels for the classical Gaussian ensembles are correlated and tend to repel each other, such that small spacing are unlikely. The corresponding spacing distributions are very similar to those derived by Wigner for random-matrix ensembles of dimension $2\times2$~\cite{meh91}. The Wigner distribution for the spacings between unfolded eigenvalues of a GOE spectrum is
\begin{equation}
P(s)=\frac{\pi}{2}\exp\left(-\frac{\pi}{4}s^2\right).
\label{C6}
\end{equation}

\subsection{Number variance $\Sigma^2$ and $\Delta_3$ statistics}

In RMT, one of the properties of the energy spectra due to the correlations among the energy level spacings is the \emph{spectral rigidity}. It is said that a spectrum is \emph{rigid} if the fluctuation in the number of levels found in an energy interval of given length, around its average, is very small. On the other hand, if the spacings between levels are not correlated it is said then that the spectrum is \emph{soft}. One of the spectral fluctuation measures that quantify the spectral rigidity is the number variance, $\Sigma^2$. It is defined as the average variance of the level number, $n(L)$, calculated in an unfolded energy interval of length $L$. In this way, we have that $\Sigma^2$ is given by
\begin{equation}
\Sigma^2(L)=\langle n(L,x)^2\rangle-\langle n(L,x)\rangle^2,
\label{C7}
\end{equation}
where $n(L,x)$ counts the number of levels in the interval $[x,x+L]$ and $\langle\cdot\rangle$ indicates an average over the spectrum. For a sequence of uncorrelated levels, as in a Poisson spectrum, we have
\begin{equation}
\Sigma^2(L)=L.
\label{C8}
\end{equation}
Thus, for a soft spectrum, the variance of the number of levels within an energy window, increases linearly with the size of the window. For the classical Gaussian ensembles, and for large values of $L$, the increase of $\Sigma^2$ with $L$ is less than linear. In particular, for GOE the number variance depends logarithmically on $L$, in the following way
\begin{equation}
\Sigma^2(L)=\frac{2}{\pi^2}\left[\log(2\pi L)+\Gamma+1-\frac{\pi^2}{8}\right],
\label{C9}
\end{equation}
where $\Gamma$ is the Euler's constant. Another measure of the rigidity of a spectrum in RMT is the $\Delta_3$ statistics introduced by Dyson and Mehta~\cite{meh91}. For an unfolded eigenvalue sequence of length $L$, $[x,x+L]$, it is defined as the average deviation of the accumulated level density function, $\mathcal{N}(x)=\int_{-\infty}^x\rho(x')dx'$, from the best straight line adjusted by the least squares method,
\begin{equation}
\Delta_3(L)=\frac{1}{L}\left\langle \underset{A,B}{\min}\int_x^{x+L}dx'\hspace{0.5ex}[\mathcal{N}(x')-Ax'-B]^2\right\rangle
\label{C10}
\end{equation}
where $\langle\cdot\rangle$ denotes a spectral average. For a given $L$, the smaller the value of $\Delta_3$ the stronger the rigidity is. For uncorrelated spectra, i.e., without level repulsion, as in the case of the Poisson ensemble, we have
\begin{equation}
\Delta_3(L)=L/15,
\label{C11}
\end{equation}
while for GOE, the ensemble average of $\Delta_3$ depends logarithmically of $L$ for large values of $L$, such that
\begin{equation}
\Delta_3(L)=\frac{1}{\pi^2}\left[\log(2\pi L)+\Gamma-\frac{5}{4}-\frac{\pi^2}{8}\right].
\label{C12}
\end{equation}

\section{Results and discussion}
\label{SecIV}

Now, we apply all spectral measures described above to the spectra of the adjacency matrices of three representations of ER networks. We call {\it fully random-weighted ER networks} to the first of the three ER networks we study here. This random network ensemble has already been studied in Refs.~\cite{jac01,Torres18,net01} and is constructed as follows. Starting with the standard ER network, we add self-edges and further consider all edges to have random strengths. The sparsity parameter $\alpha$ (average network connectivity) is defined as the fraction of the $N(N-1)/2$ independent non-vanishing off-diagonal adjacency matrix elements. The corresponding adjacency matrices, ${\bf A}$, come from the ensemble of $N\times N$ sparse real symmetric matrices, whose non-vanishing elements are statistically independent random variables drawn from a Gaussian distribution,
\begin{equation}
P(A_{ij})=\frac{1}{\sqrt{2 \pi \sigma_{ij}^2}} \exp \left( - \frac{A_{ij}^2}{2 \sigma_{ij}^2} \right),
\label{B2}
\end{equation}
with $\sigma_{ij}=1+\delta_{ij}$. According to this definition, Poisson statistics is recovered in the limit case when the vertices in the network are isolated $\alpha=0$ (maximum sparsity). GOE statistics is recovered in the case when the network is fully connected $\alpha=1$ (null sparsity). Although this ensemble was already discussed in Ref.~\cite{Torres18}, there, the $\alpha$ values were chosen in such a way that the crossover in the long-range correlations of the spectral fluctuations, between the GOE and Poisson limits, was clearly visualized in the scree diagrams. The latter, as it is shown there, does not allow to see in detail the transition in the short-range correlations quantified by the NNSD and the $P(\tilde{r})$ distribution.

\begin{figure}
\begin{center}
\begin{minipage}{0.77\linewidth}
\centering
\includegraphics[width=\linewidth]{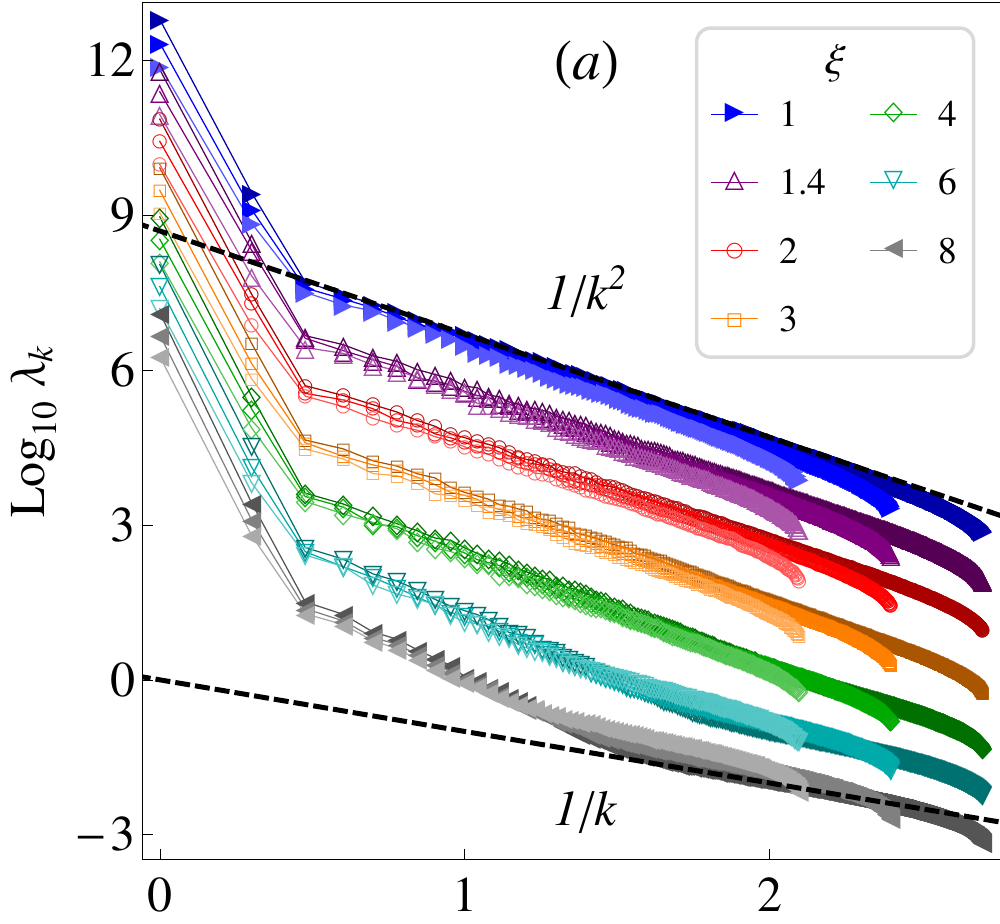}
\end{minipage}
\begin{minipage}{0.77\linewidth}
\centering
\includegraphics[width=\linewidth]{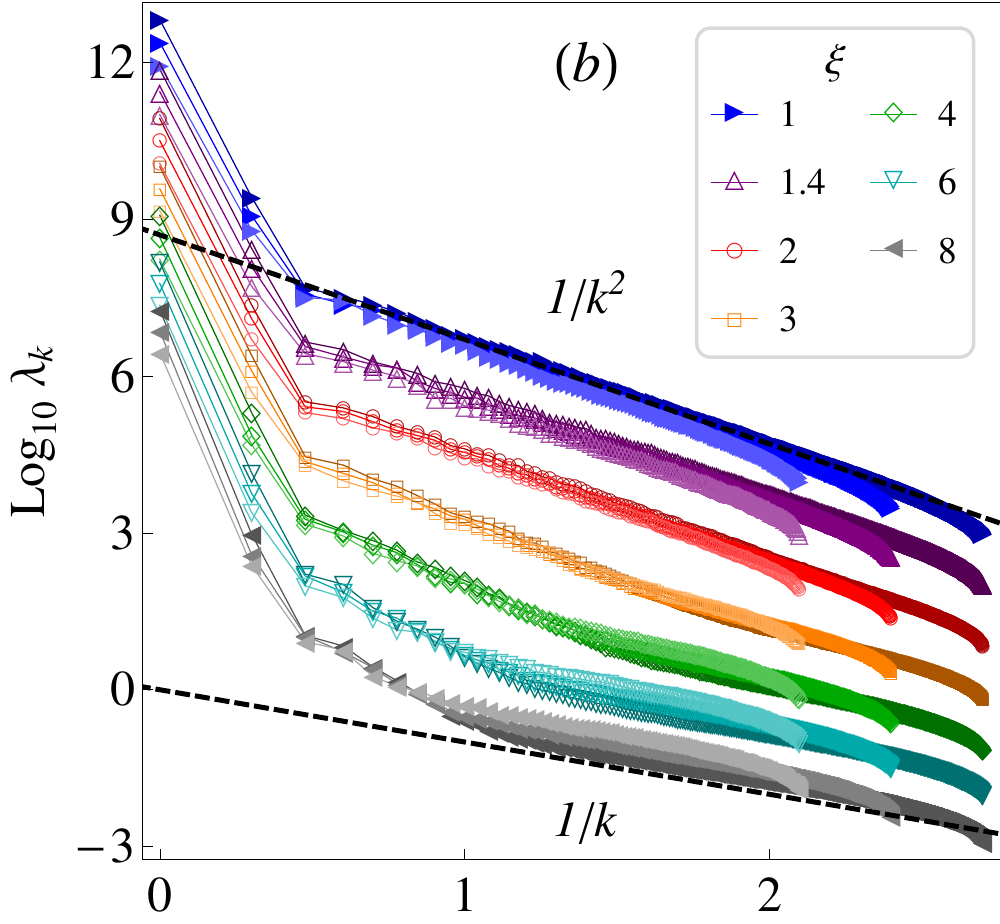}
\end{minipage}
\begin{minipage}{0.77\linewidth}
\centering
\includegraphics[width=\linewidth]{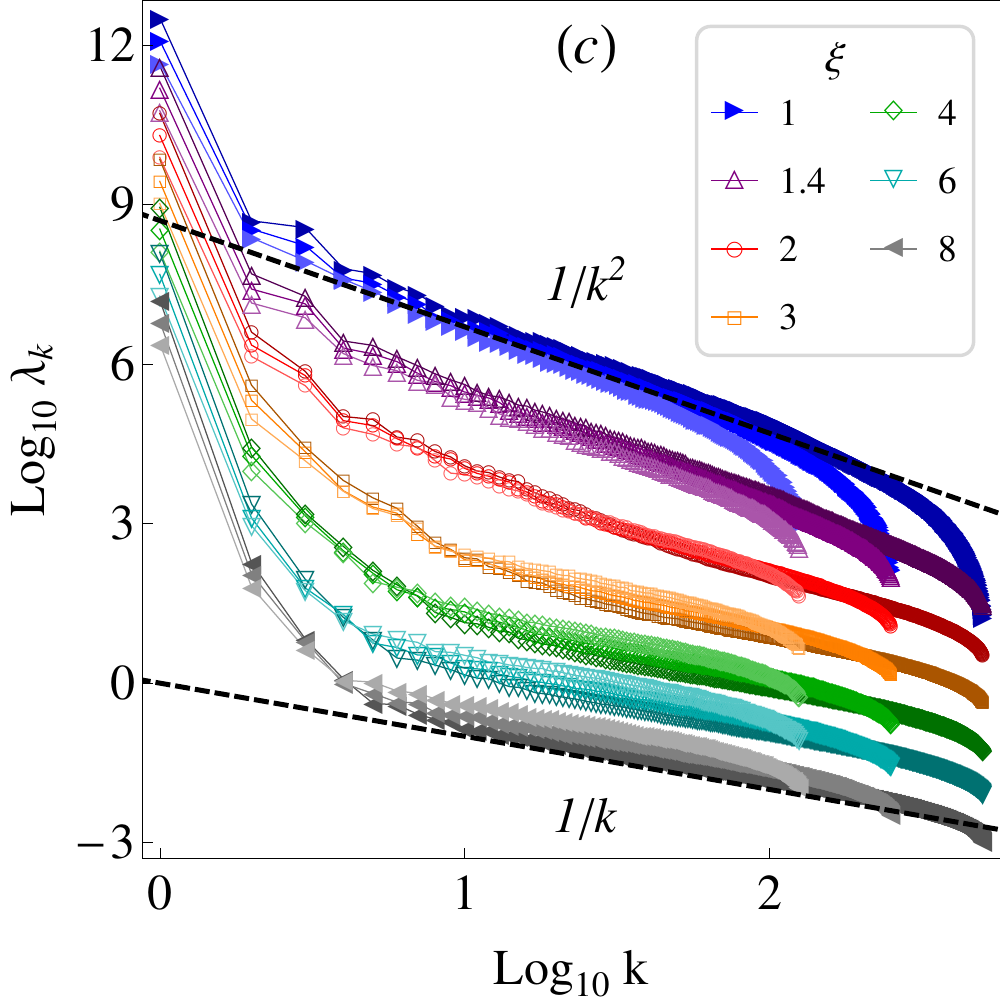}
\end{minipage}
\end{center}
\caption{Scree diagrams of ordered partial variances $\lambda_k$ for ensembles of $M=125$ eigenspectra of dimensions $N=500$ (short ones), $M=250$ eigenspectra of dimension $N=1000$ (medium size ones), and $M=500$ eigenspectra of dimension $N=2000$ (large ones). (a) Fully random-weighted ER networks, (b) ER networks with random-weighted self-edges, and (c) standard ER networks. Regardless of the size of the eigenspectra, we can appreciate a crossover in the scree diagrams between the Poisson ($1/k^2$) and GOE ($1/k$) limits for the three ER network ensembles. The scree diagrams have been shifted vertically for comparison purposes.}
\label{comboSD}
\end{figure}
\begin{figure}
\begin{minipage}[t]{0.35\linewidth}
\centering
\textbf{(\emph{a})}
  \includegraphics[width=\linewidth]{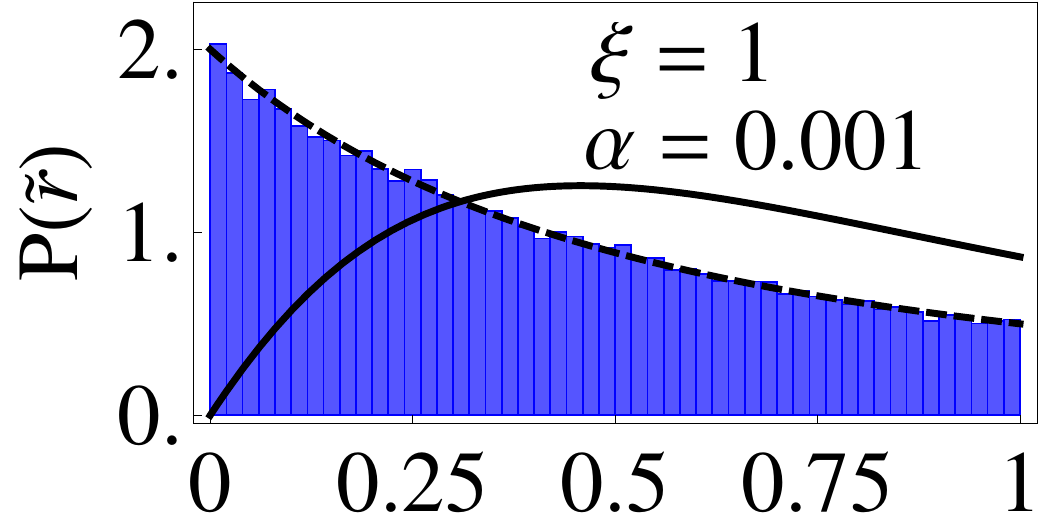}
\end{minipage}
\hfill
\begin{minipage}[t]{0.31\linewidth}
\centering
\textbf{(\emph{b})}
  \includegraphics[width=\linewidth]{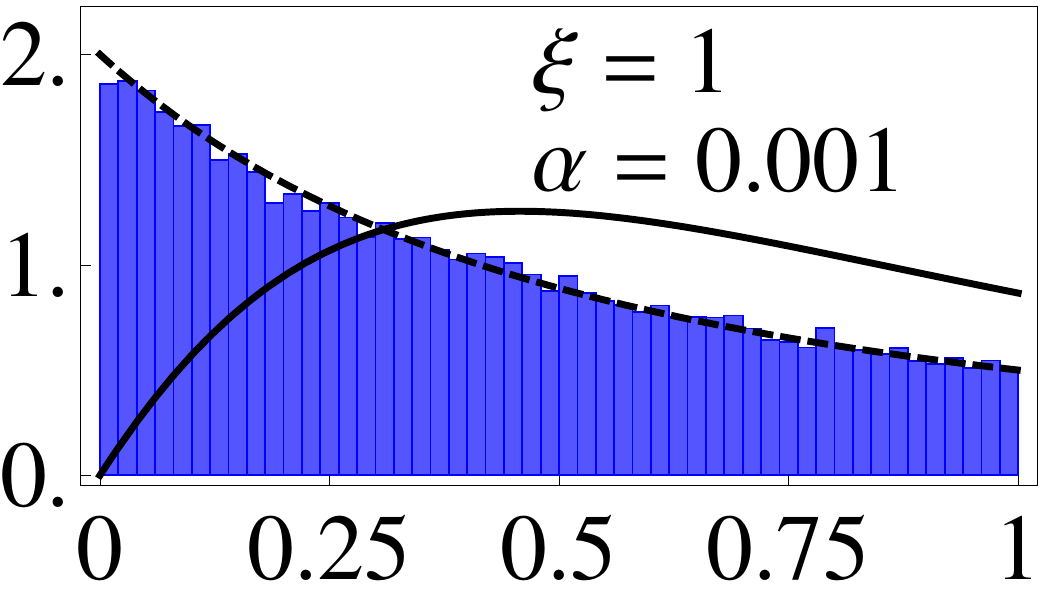}
\end{minipage}
\hfill
\begin{minipage}[t]{0.31\linewidth}
\centering
\textbf{(\emph{c})}
  \includegraphics[width=\linewidth]{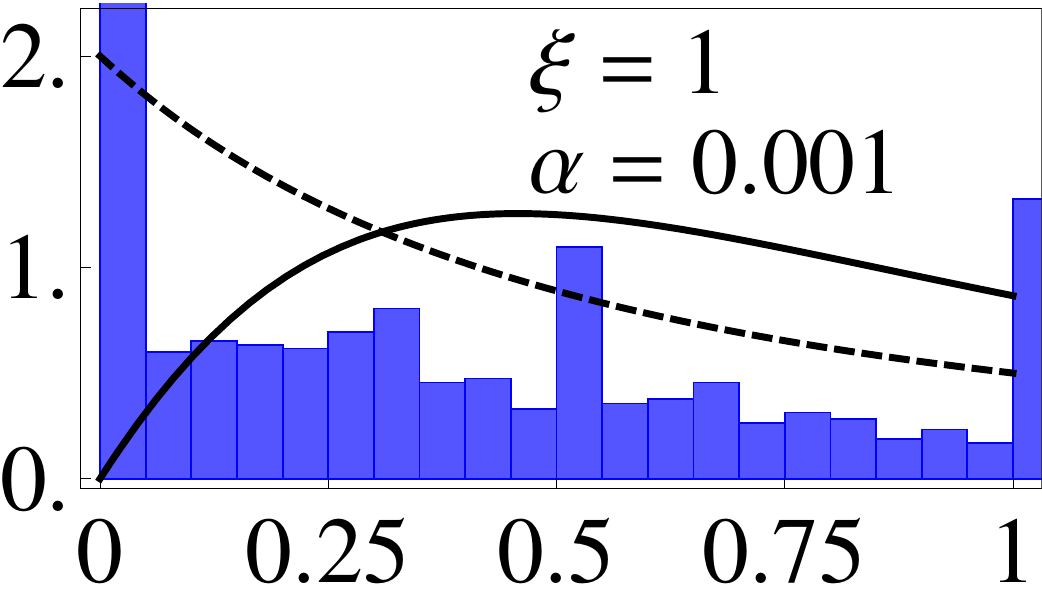}
\end{minipage}
\begin{minipage}[t]{0.35\linewidth}
\centering
  \includegraphics[width=\linewidth]{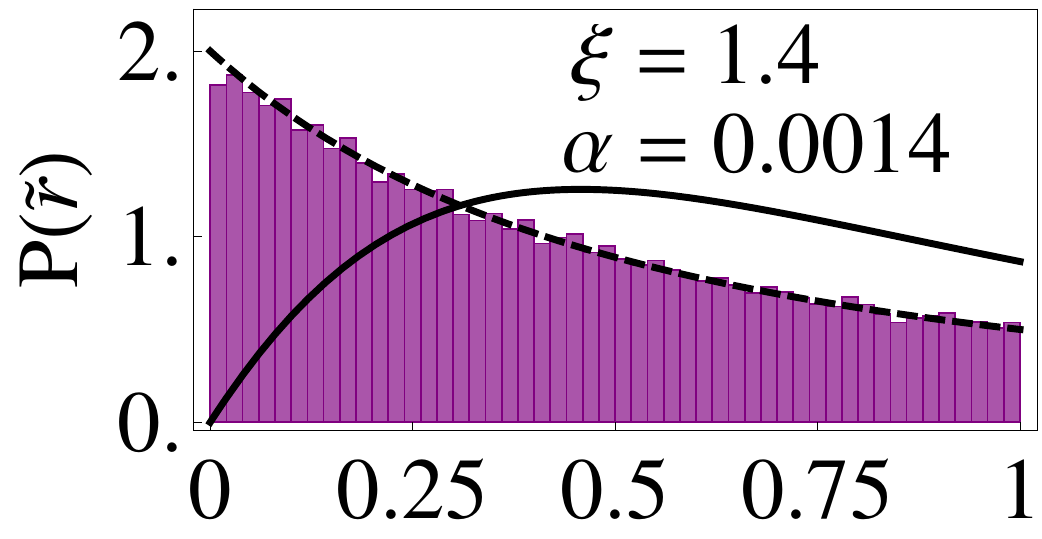}
\end{minipage}
\hfill
\begin{minipage}[t]{0.31\linewidth}
\centering
  \includegraphics[width=\linewidth]{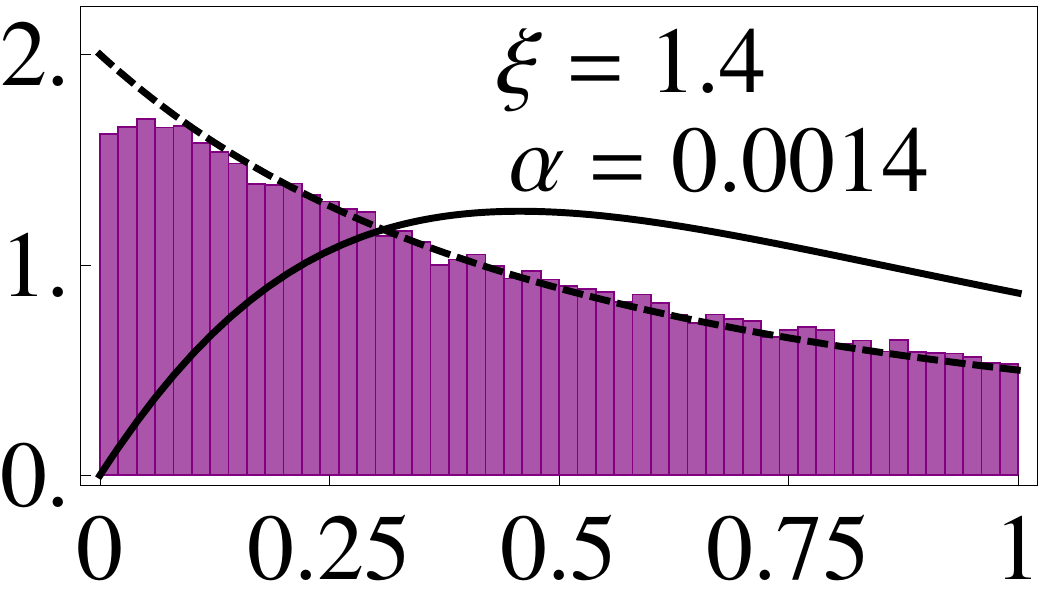}
\end{minipage}
\hfill
\begin{minipage}[t]{0.31\linewidth}
\centering
  \includegraphics[width=\linewidth]{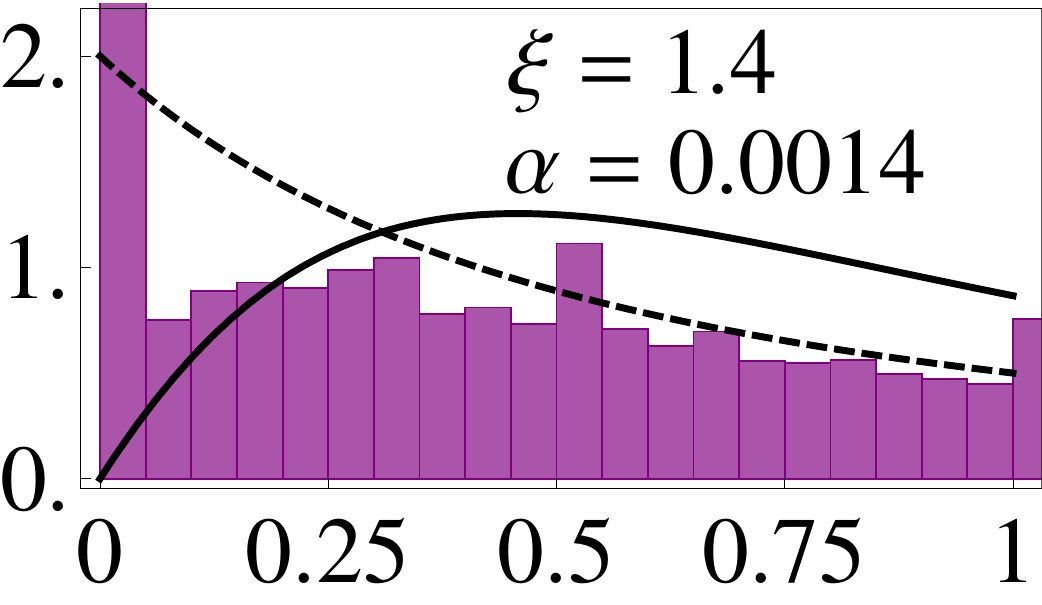}
\end{minipage}
\begin{minipage}[t]{0.35\linewidth}
\centering
  \includegraphics[width=\linewidth]{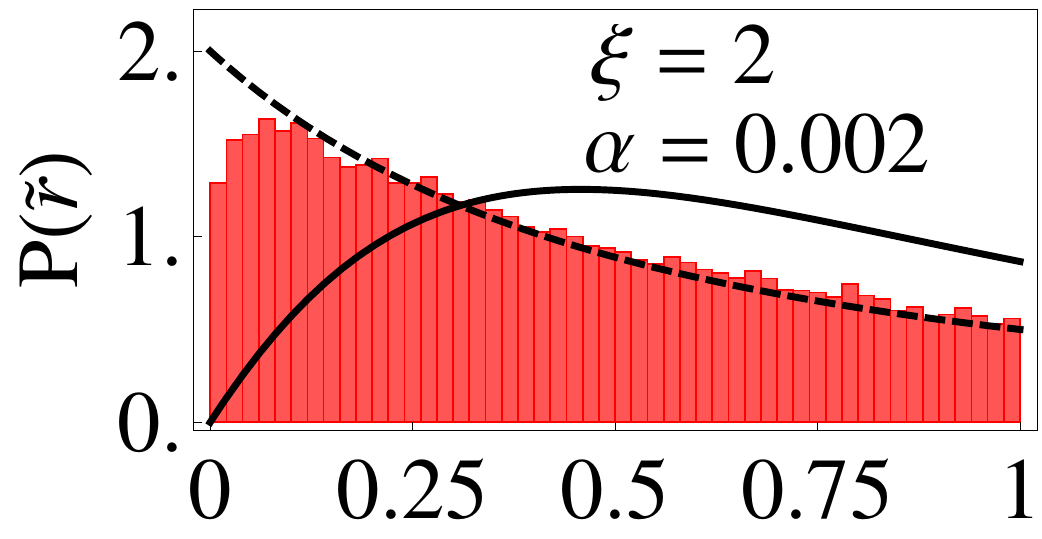}
\end{minipage}
\hfill
\begin{minipage}[t]{0.31\linewidth}
\centering
  \includegraphics[width=\linewidth]{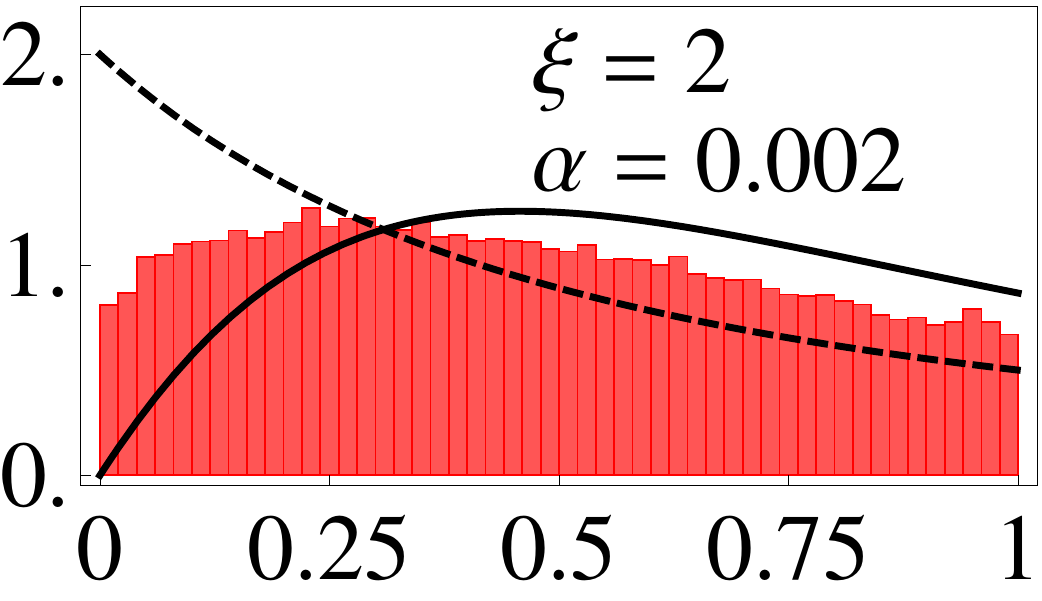}
\end{minipage}
\hfill
\begin{minipage}[t]{0.31\linewidth}
\centering
  \includegraphics[width=\linewidth]{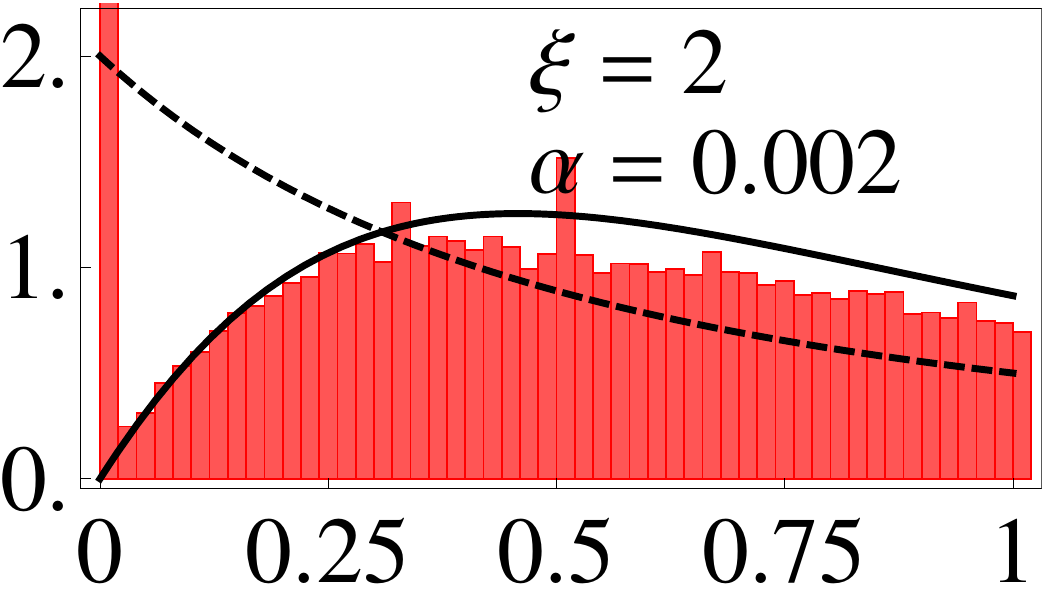}
\end{minipage}
\begin{minipage}[t]{0.35\linewidth}
\centering
  \includegraphics[width=\linewidth]{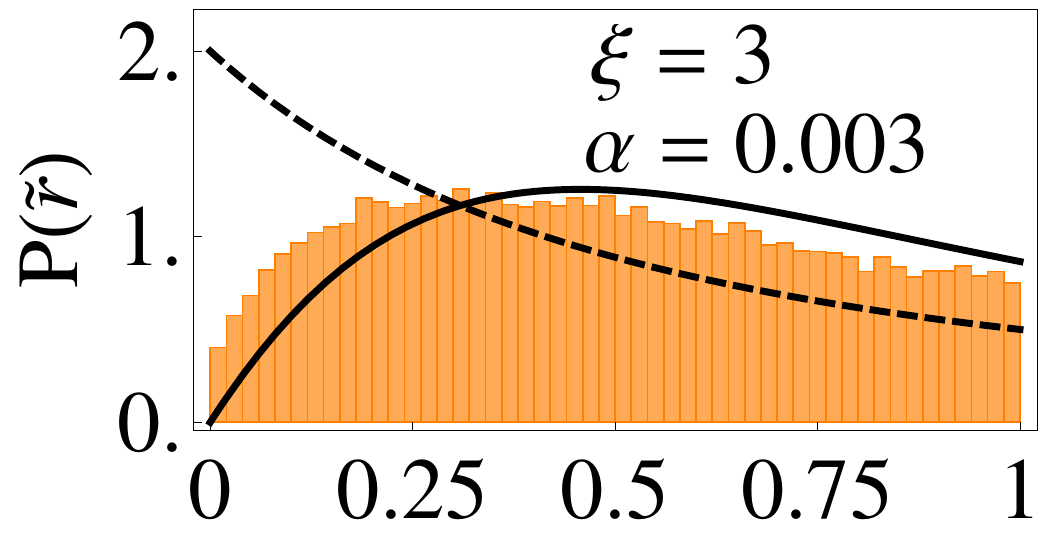}
\end{minipage}
\hfill
\begin{minipage}[t]{0.31\linewidth}
\centering
  \includegraphics[width=\linewidth]{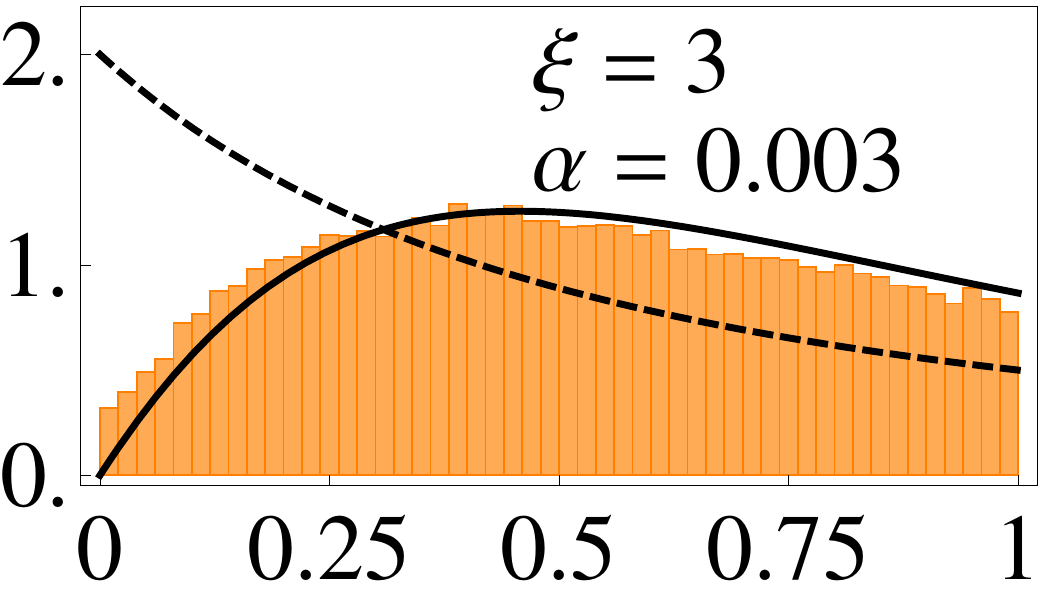}
\end{minipage}
\hfill
\begin{minipage}[t]{0.31\linewidth}
\centering
  \includegraphics[width=\linewidth]{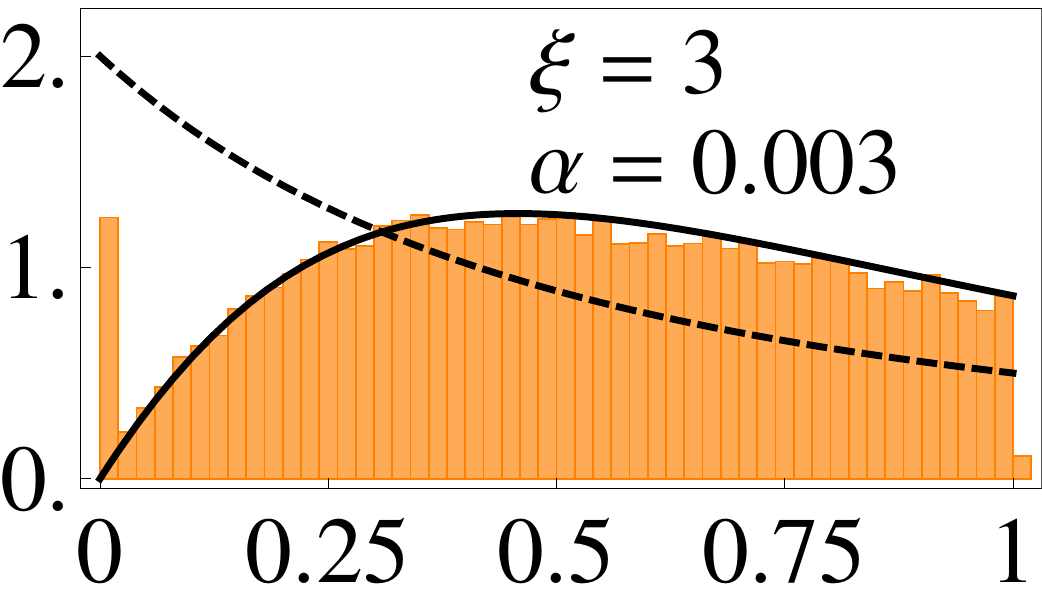}
\end{minipage}
\begin{minipage}[t]{0.35\linewidth}
\centering
  \includegraphics[width=\linewidth]{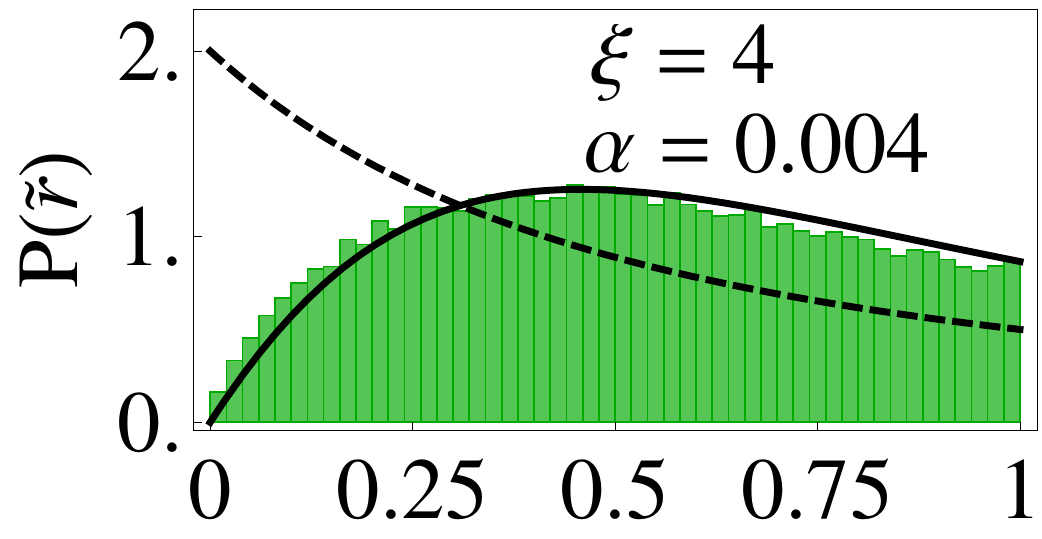}
\end{minipage}
\hfill
\begin{minipage}[t]{0.31\linewidth}
\centering
  \includegraphics[width=\linewidth]{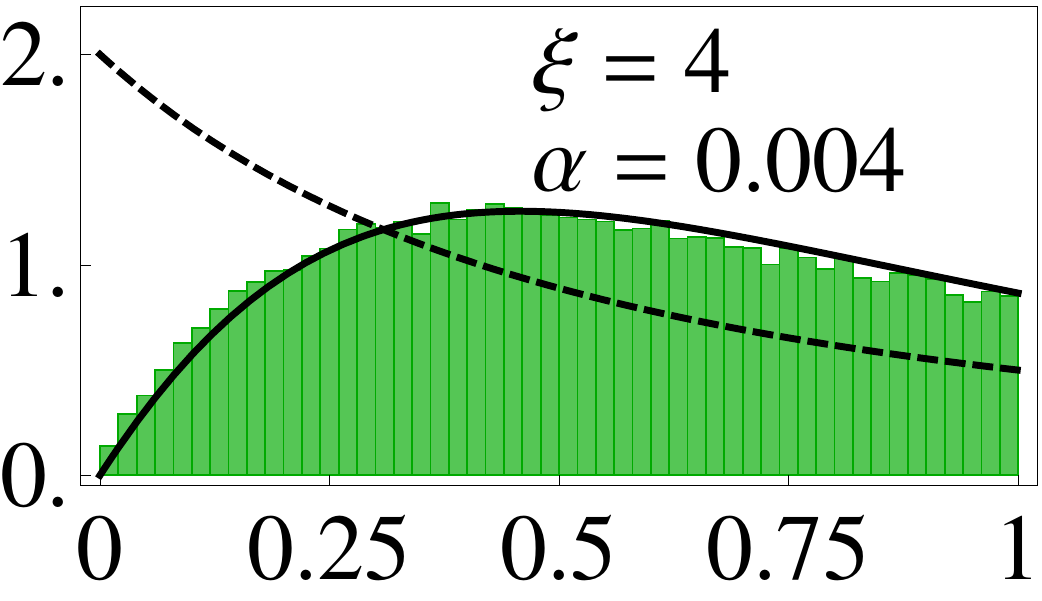}
\end{minipage}
\hfill
\begin{minipage}[t]{0.31\linewidth}
\centering
  \includegraphics[width=\linewidth]{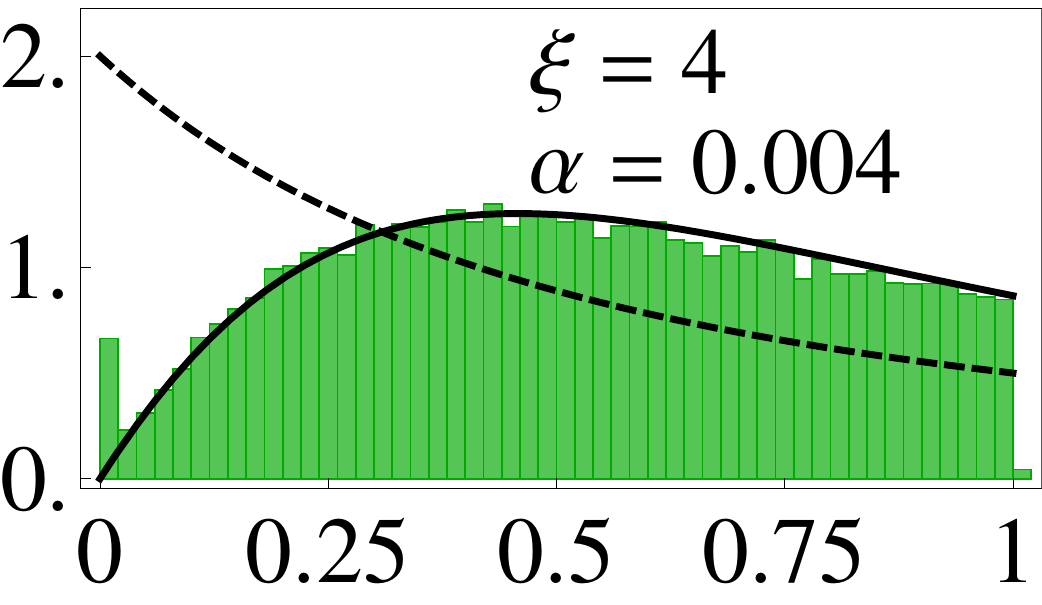}
\end{minipage}
\begin{minipage}[t]{0.35\linewidth}
\centering
  \includegraphics[width=\linewidth]{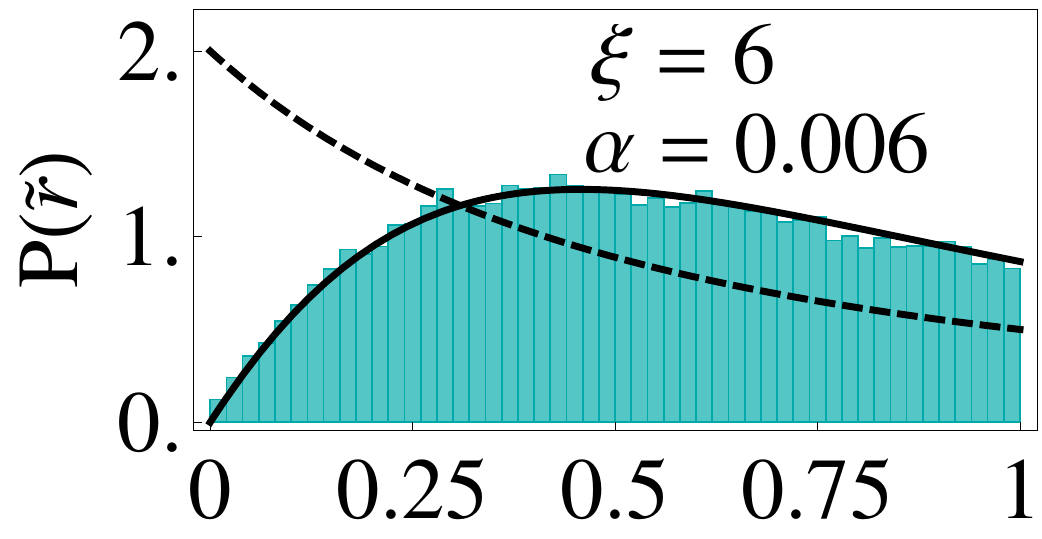}
\end{minipage}
\hfill
\begin{minipage}[t]{0.31\linewidth}
\centering
  \includegraphics[width=\linewidth]{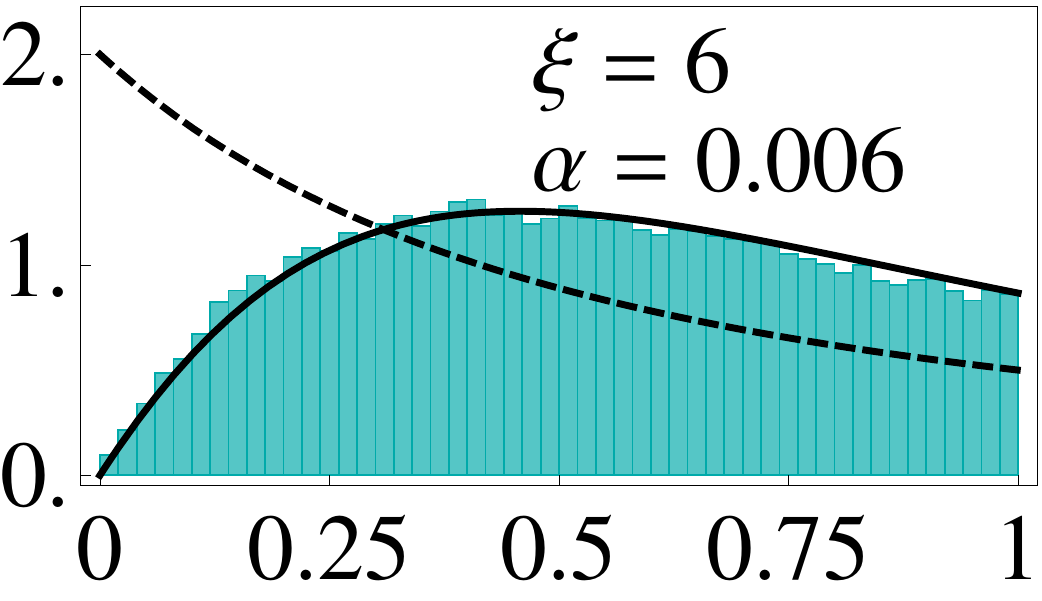}
\end{minipage}
\hfill
\begin{minipage}[t]{0.31\linewidth}
\centering
  \includegraphics[width=\linewidth]{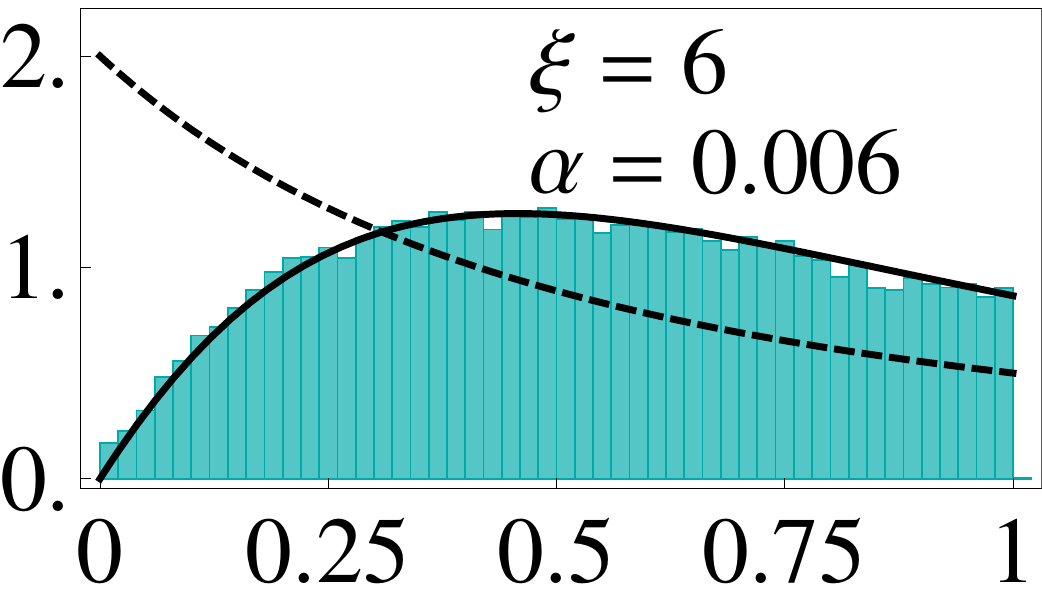}
\end{minipage}
\begin{minipage}[t]{0.35\linewidth}
\centering
  \includegraphics[width=\linewidth]{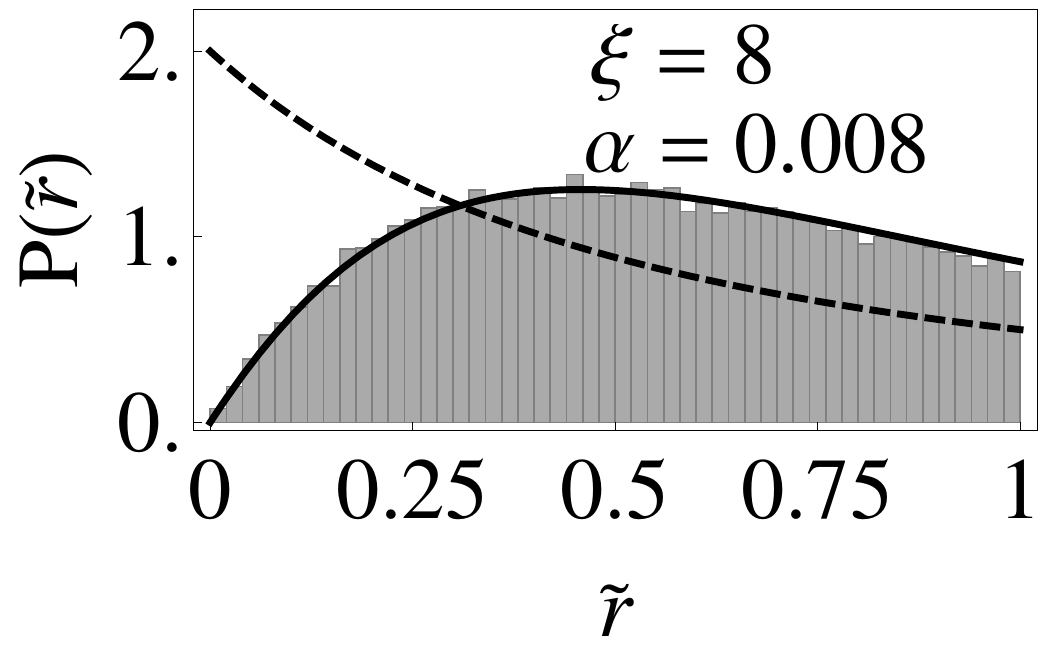}
\end{minipage}
\hfill
\begin{minipage}[t]{0.31\linewidth}
\centering
  \includegraphics[width=\linewidth]{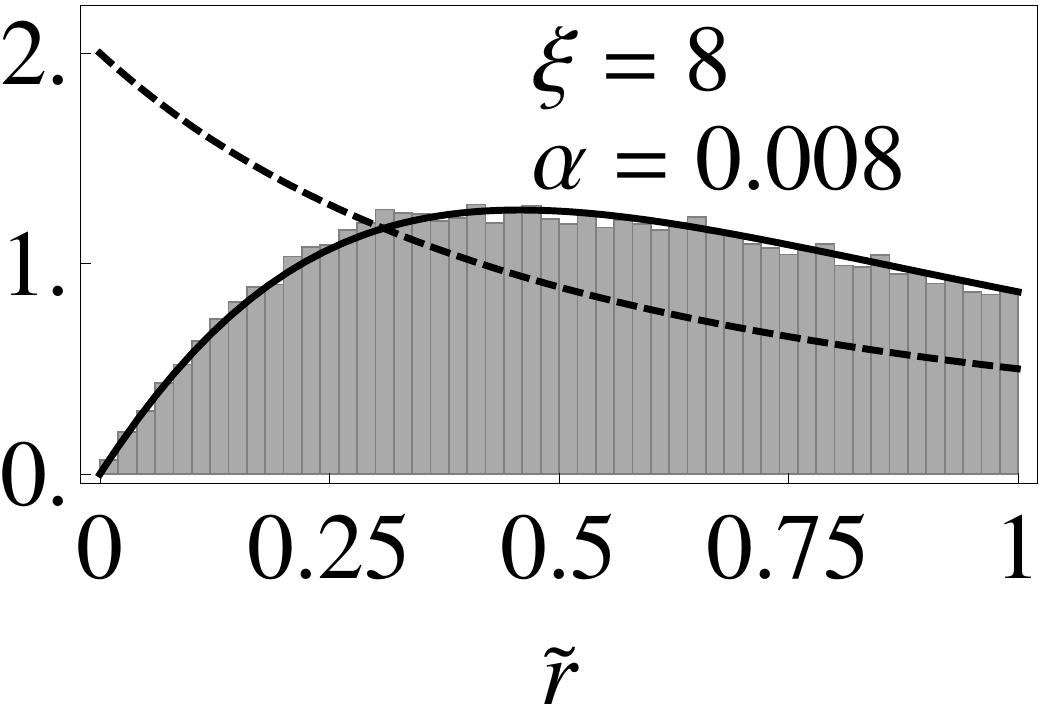}
\end{minipage}
\hfill
\begin{minipage}[t]{0.31\linewidth}
\centering
  \includegraphics[width=\linewidth]{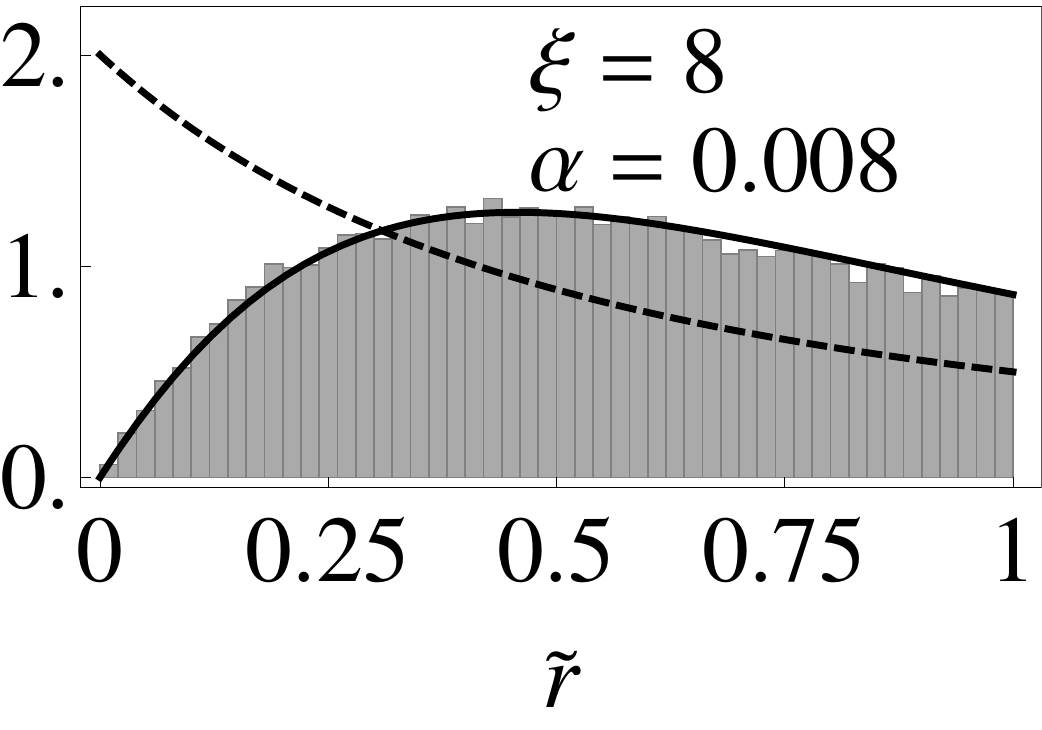}
\end{minipage}
\caption{Distribution $P(\tilde{r})$ for ensembles of $M=250$ eigenspectra of dimension $N=1000$, corresponding to the three ER networks reported in Fig.~\ref{comboSD}: (a) fully random-weighted ER networks, (b) ER networks with random-weighted self-edges, and (c) standard ER networks. As we can see, a transition from Poisson to GOE statistics is obtained as the average degree $\xi$, or equivalently, the average network connectivity $\alpha$ increases. Dashed and full lines correspond to the Poisson and GOE limits, respectively.}
\label{comboPr}
\end{figure}

The second ensemble of ER networks are {\it ER networks with random-weighted self-edges}. We construct this random network model by adding self-edges with random strengths to standard ER networks. The main diagonal of the corresponding adjacency matrices acquire statistically independent random variables, drawn from a Gaussian distribution with zero mean and variance one. With this construction, a clear Poisson to Wigner-Dyson transition in the form of $P(s)$, when $\alpha$ moves from zero to one, was observed in~\cite{net01}.

\begin{figure}
\begin{center}
\begin{minipage}{0.75\linewidth}
\centering
\includegraphics[width=\linewidth]{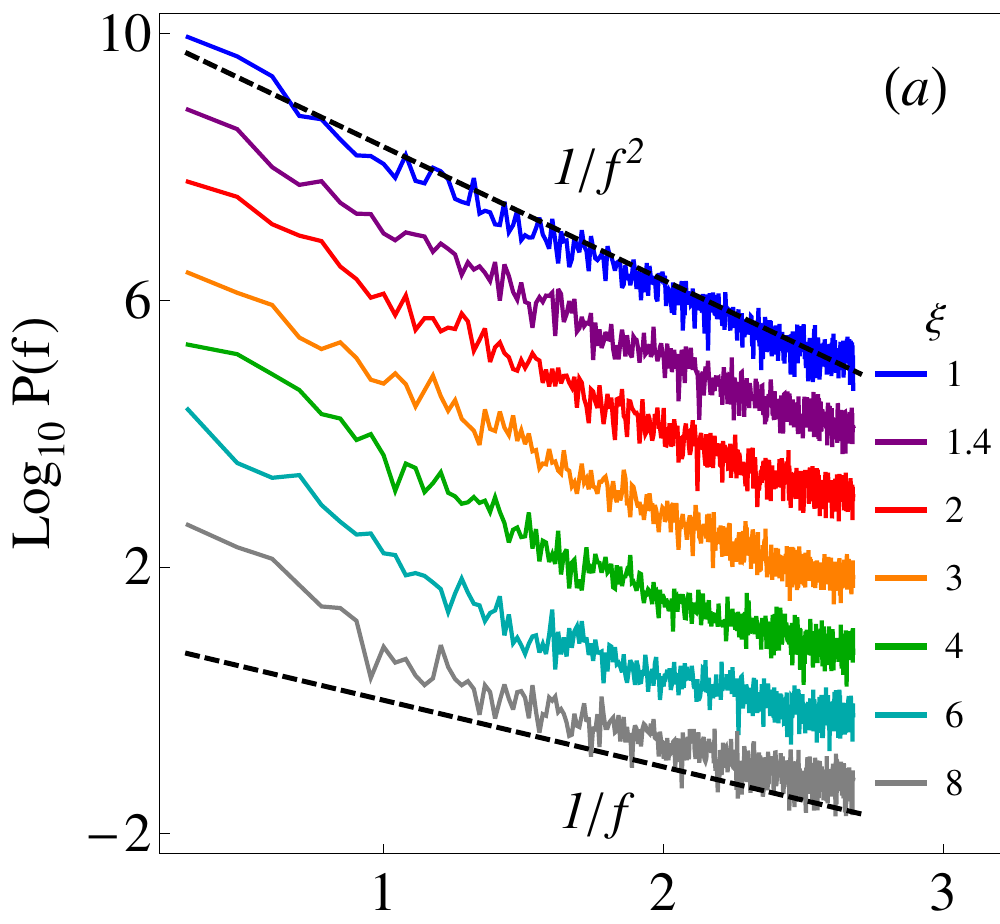}
\end{minipage}
\begin{minipage}{0.75\linewidth}
\centering
\includegraphics[width=\linewidth]{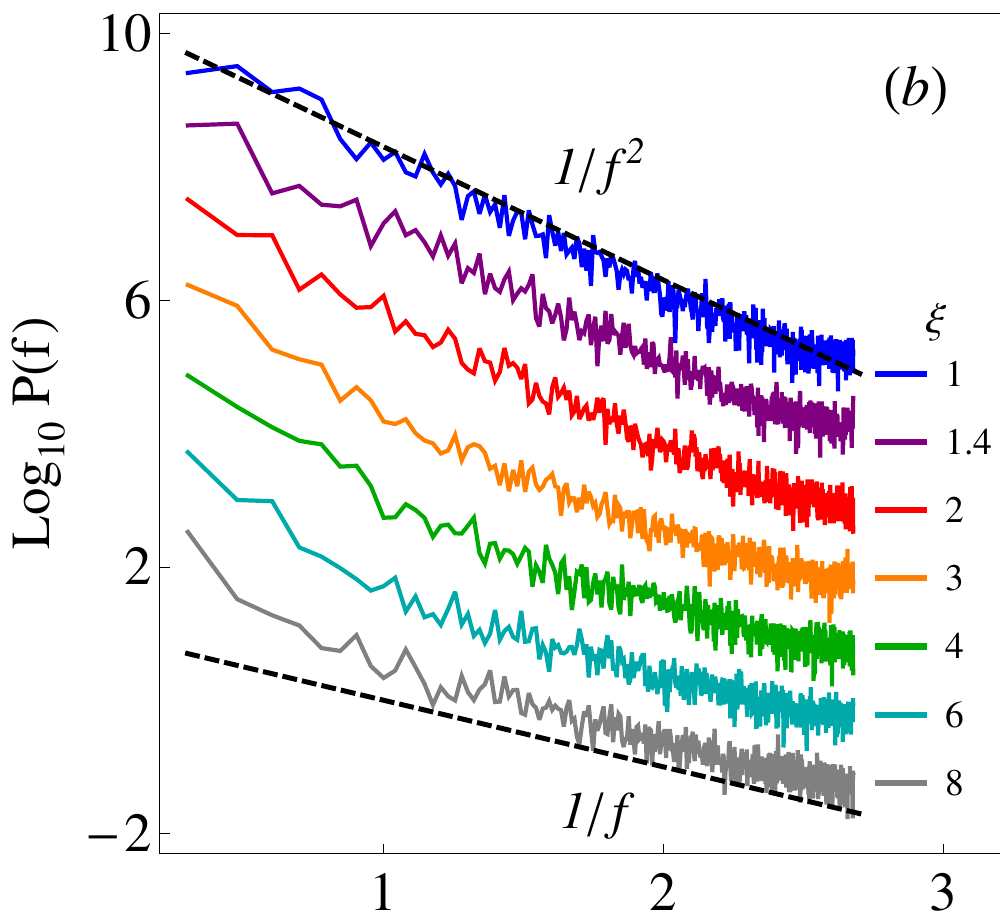}
\end{minipage}
\begin{minipage}{0.75\linewidth}
\centering
\includegraphics[width=\linewidth]{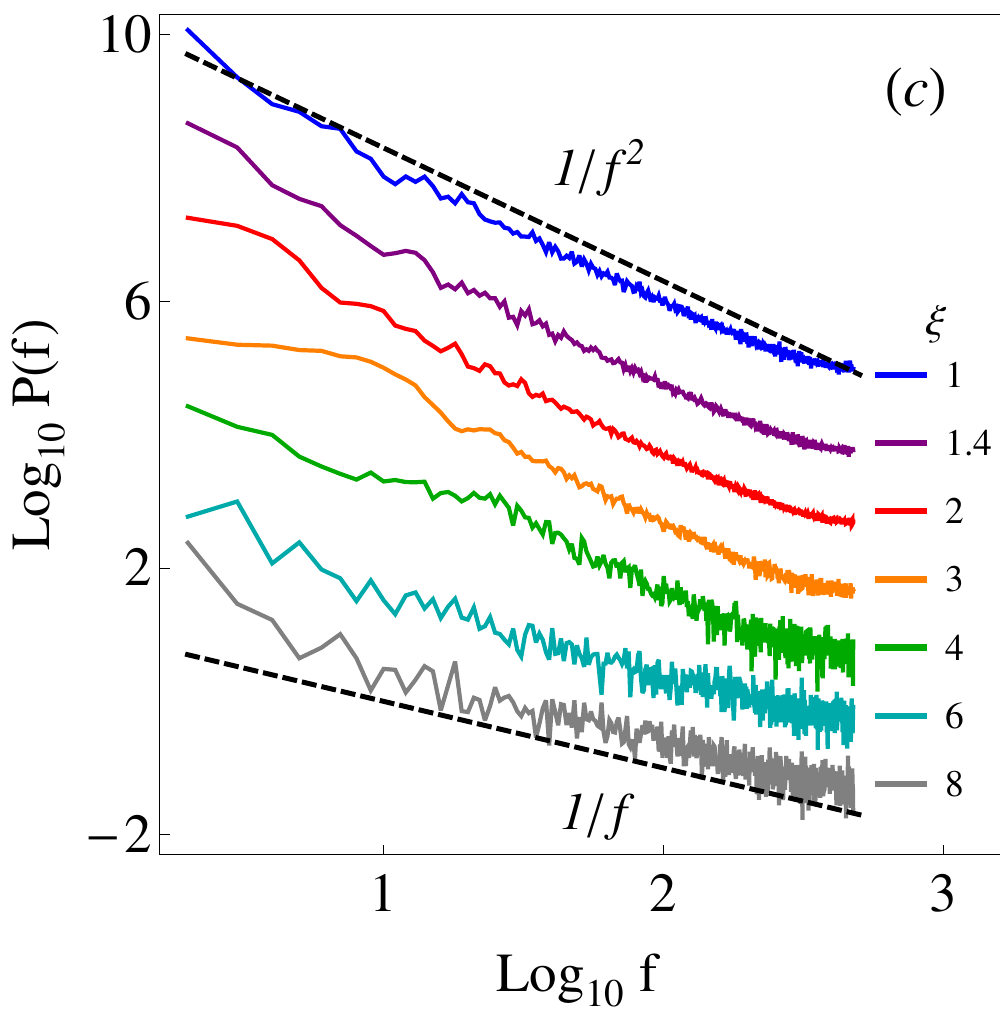}
\end{minipage}
\end{center}
\caption{Power spectra, $P(f)$, obtained from the $\delta_n$ statistics, for ensembles of $M=250$ eigenspectra of dimension $N=1000$, after performing a data-adaptive unfolding, for (a) fully random-weighted ER networks, (b) ER networks with random-weighted self-edges, and (c) standard ER networks. The values of the average degree $\xi$, and therefore the values of the average network connectivity $\alpha$, are the same in all cases. As in the scree diagrams presented in Fig.~\ref{comboSD}, the power spectra exhibit a crossover between Poisson ($1/f^2$) and GOE ($1/f$) limits. The results correspond to ensemble averages. The power spectra have been shifted vertically for comparison purposes.}
\label{comboPS}
\end{figure}
\begin{figure}
\begin{minipage}[t]{0.35\linewidth}
\centering
\textbf{(\emph{a})}
  \includegraphics[width=\linewidth]{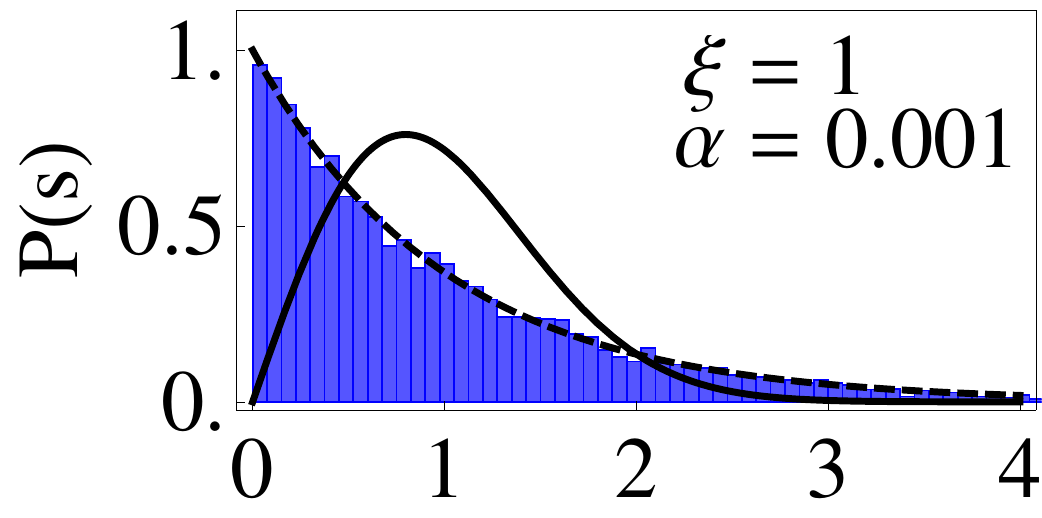}
\end{minipage}
\hfill
\begin{minipage}[t]{0.31\linewidth}
\centering
\textbf{(\emph{b})}
  \includegraphics[width=\linewidth]{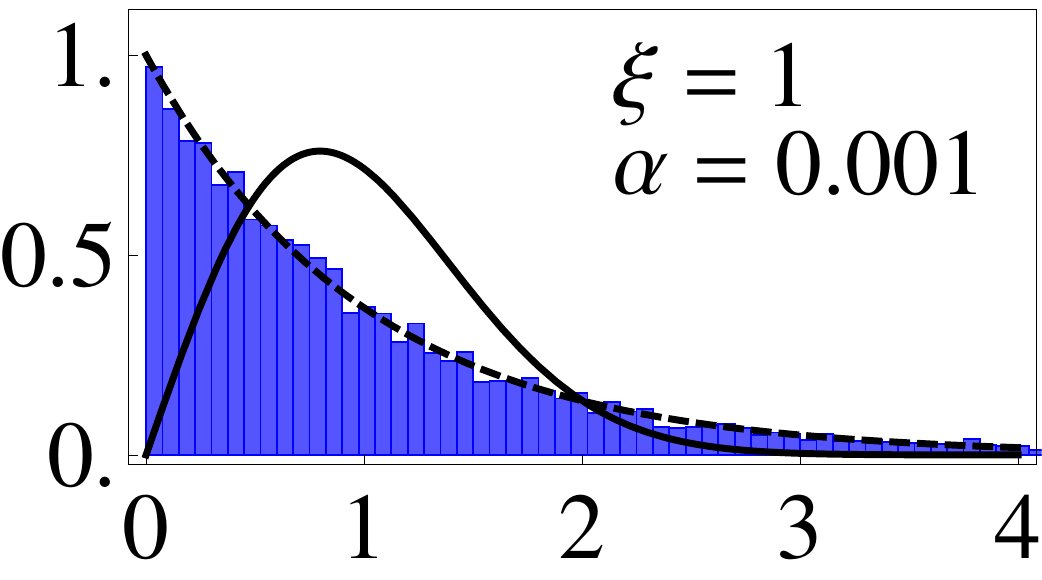}
\end{minipage}
\hfill
\begin{minipage}[t]{0.31\linewidth}
\centering
\textbf{(\emph{c})}
  \includegraphics[width=\linewidth]{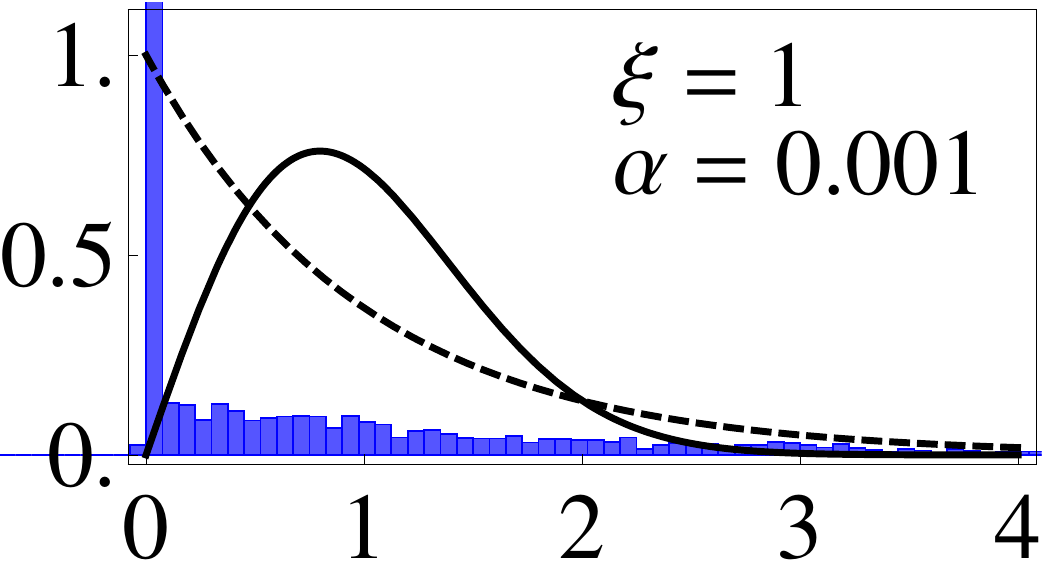}
\end{minipage}
\begin{minipage}[t]{0.35\linewidth}
\centering
  \includegraphics[width=\linewidth]{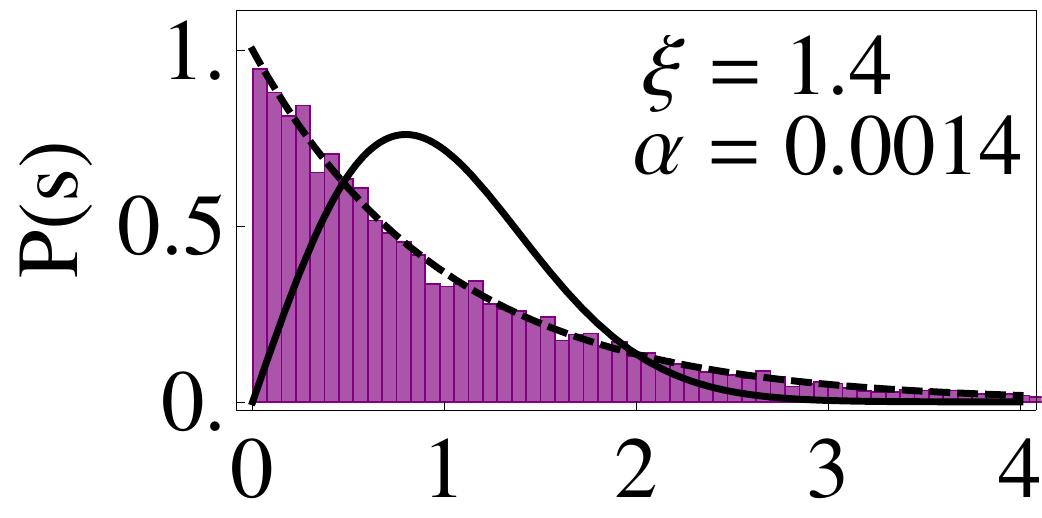}
\end{minipage}
\hfill
\begin{minipage}[t]{0.31\linewidth}
\centering
  \includegraphics[width=\linewidth]{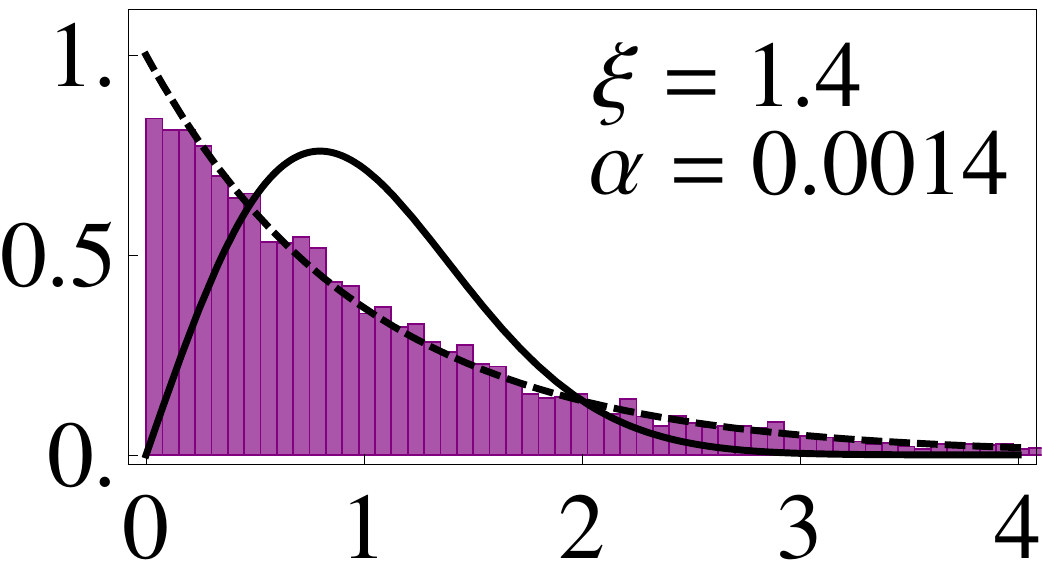}
\end{minipage}
\hfill
\begin{minipage}[t]{0.31\linewidth}
\centering
  \includegraphics[width=\linewidth]{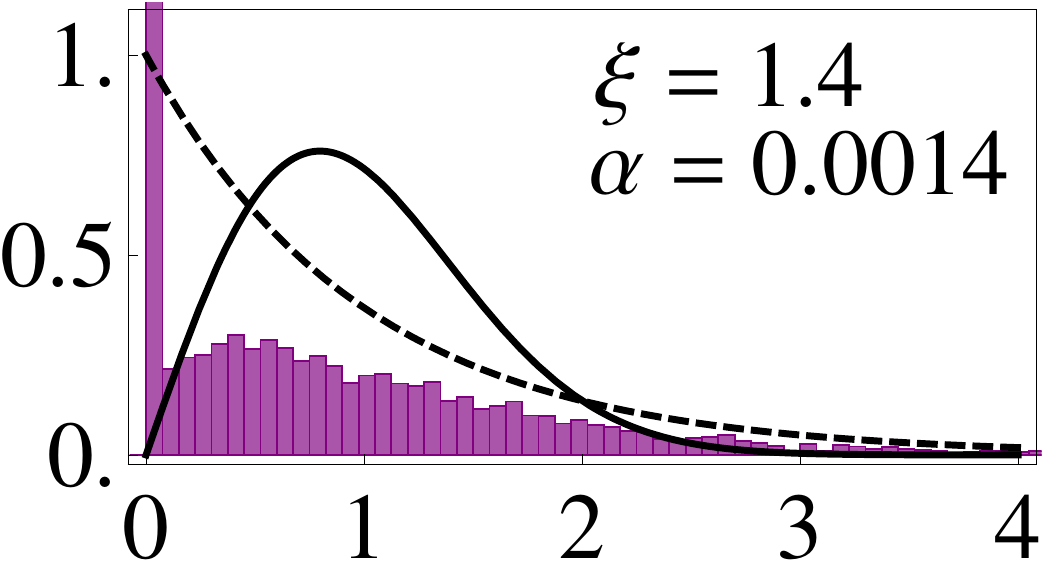}
\end{minipage}
\begin{minipage}[t]{0.35\linewidth}
\centering
  \includegraphics[width=\linewidth]{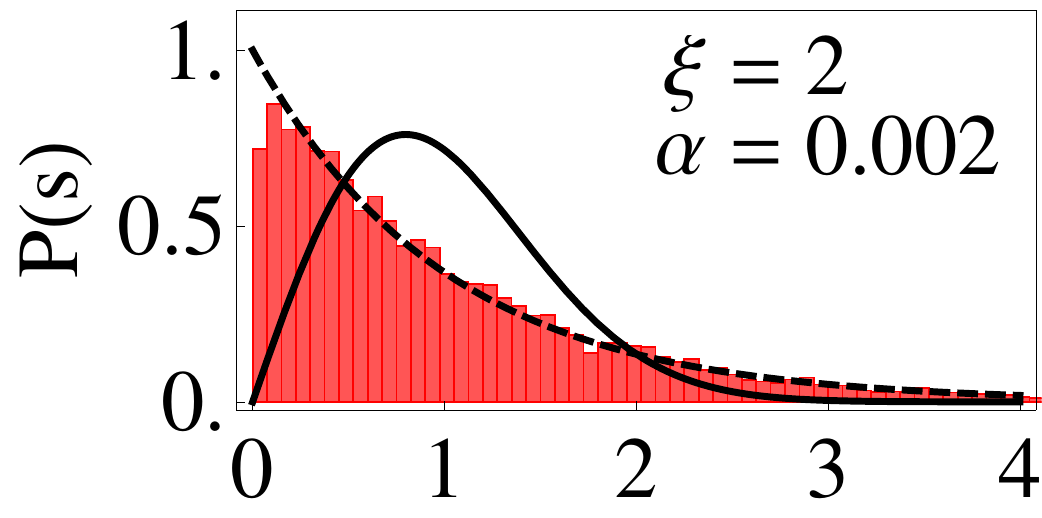}
\end{minipage}
\hfill
\begin{minipage}[t]{0.31\linewidth}
\centering
  \includegraphics[width=\linewidth]{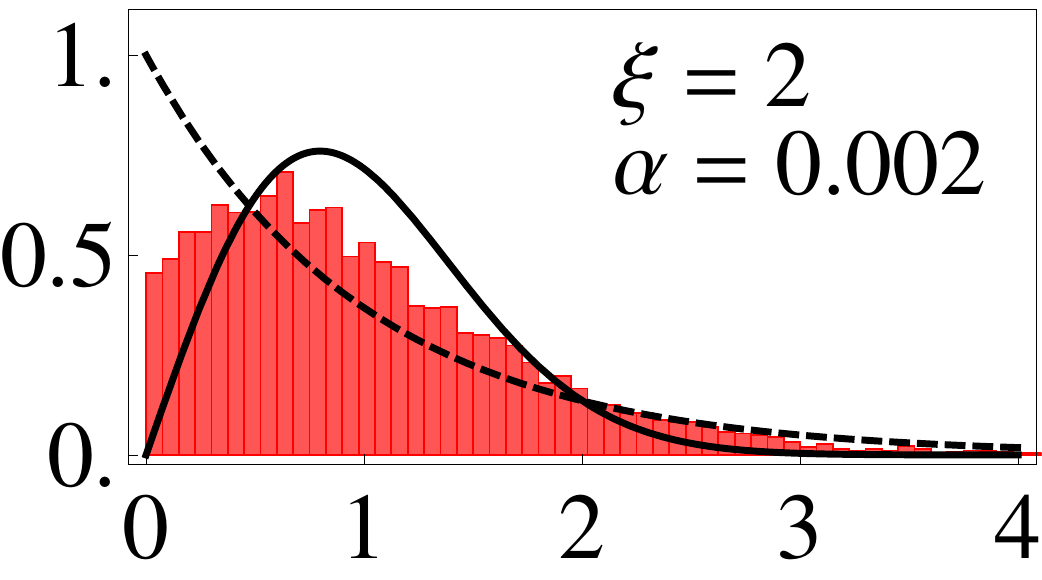}
\end{minipage}
\hfill
\begin{minipage}[t]{0.31\linewidth}
\centering
  \includegraphics[width=\linewidth]{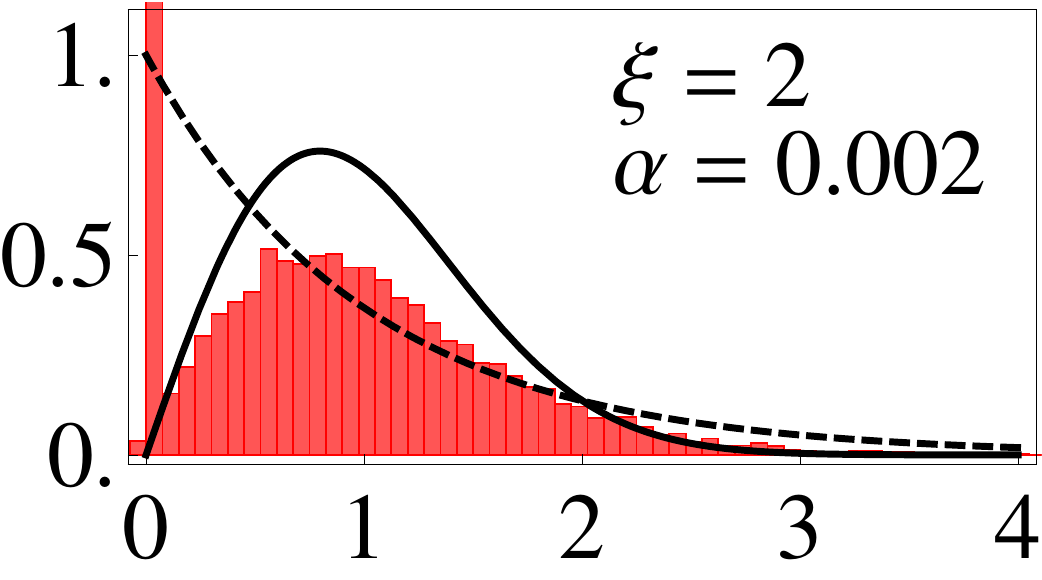}
\end{minipage}
\begin{minipage}[t]{0.35\linewidth}
\centering
  \includegraphics[width=\linewidth]{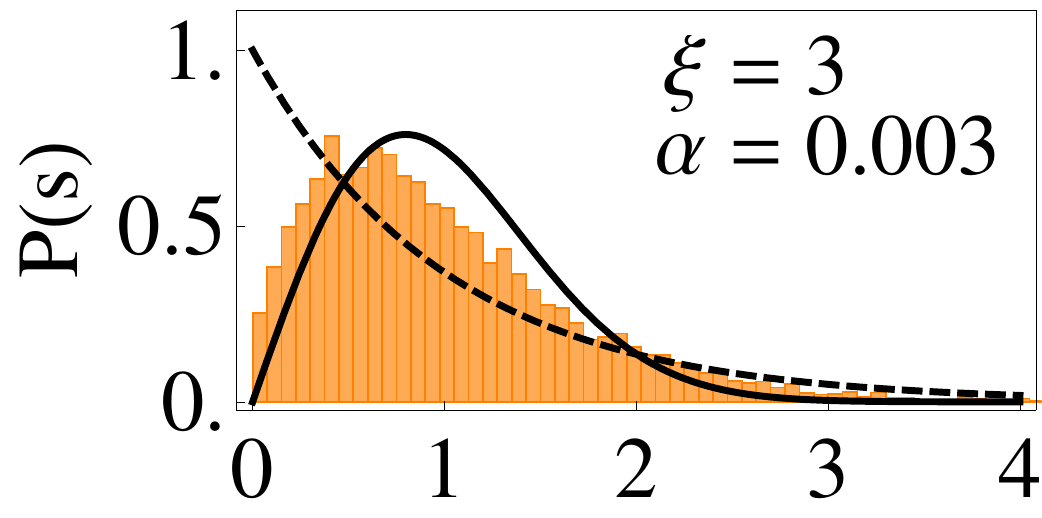}
\end{minipage}
\hfill
\begin{minipage}[t]{0.31\linewidth}
\centering
  \includegraphics[width=\linewidth]{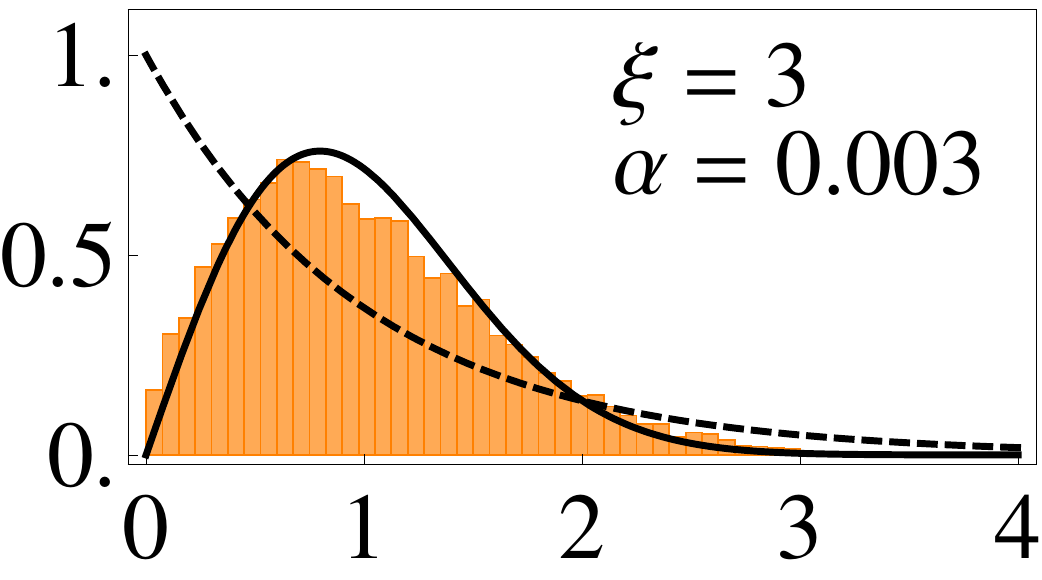}
\end{minipage}
\hfill
\begin{minipage}[t]{0.31\linewidth}
\centering
  \includegraphics[width=\linewidth]{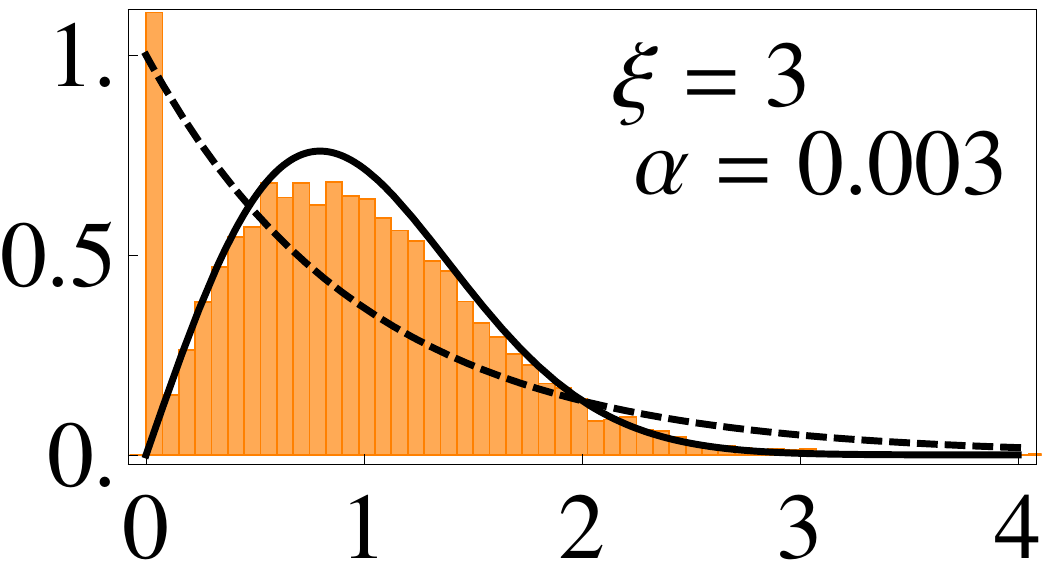}
\end{minipage}
\begin{minipage}[t]{0.35\linewidth}
\centering
  \includegraphics[width=\linewidth]{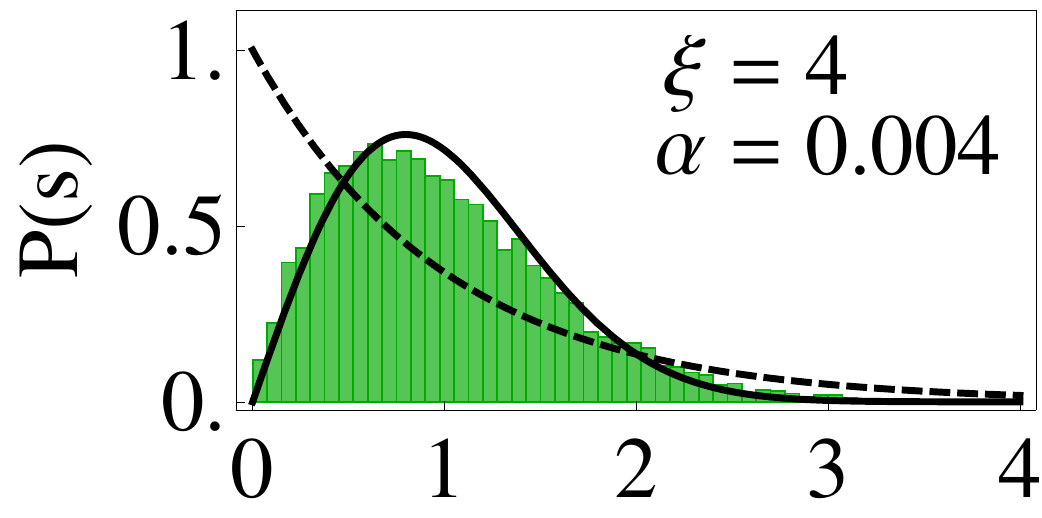}
\end{minipage}
\hfill
\begin{minipage}[t]{0.31\linewidth}
\centering
  \includegraphics[width=\linewidth]{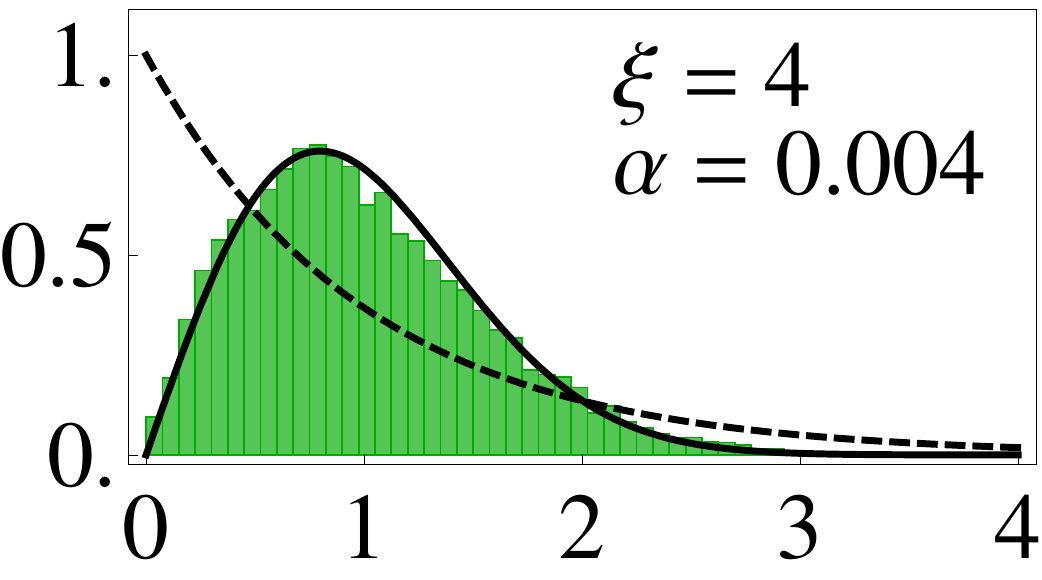}
\end{minipage}
\hfill
\begin{minipage}[t]{0.31\linewidth}
\centering
  \includegraphics[width=\linewidth]{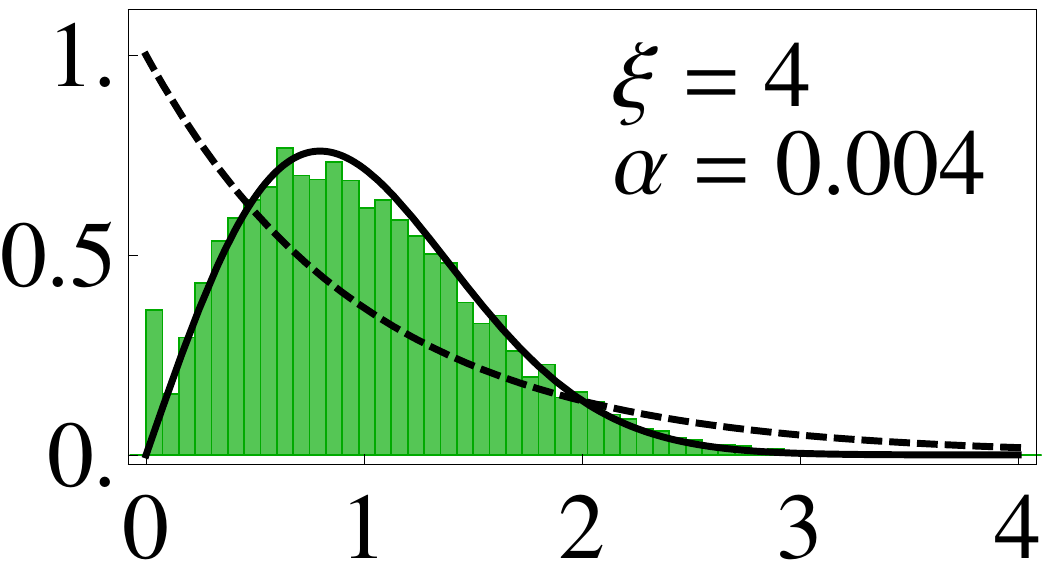}
\end{minipage}
\begin{minipage}[t]{0.35\linewidth}
\centering
  \includegraphics[width=\linewidth]{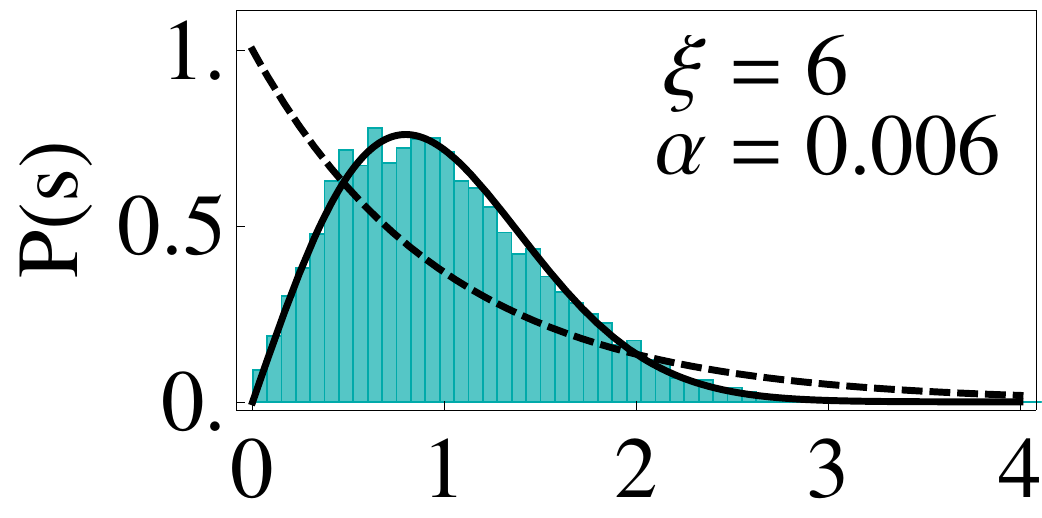}
\end{minipage}
\hfill
\begin{minipage}[t]{0.31\linewidth}
\centering
  \includegraphics[width=\linewidth]{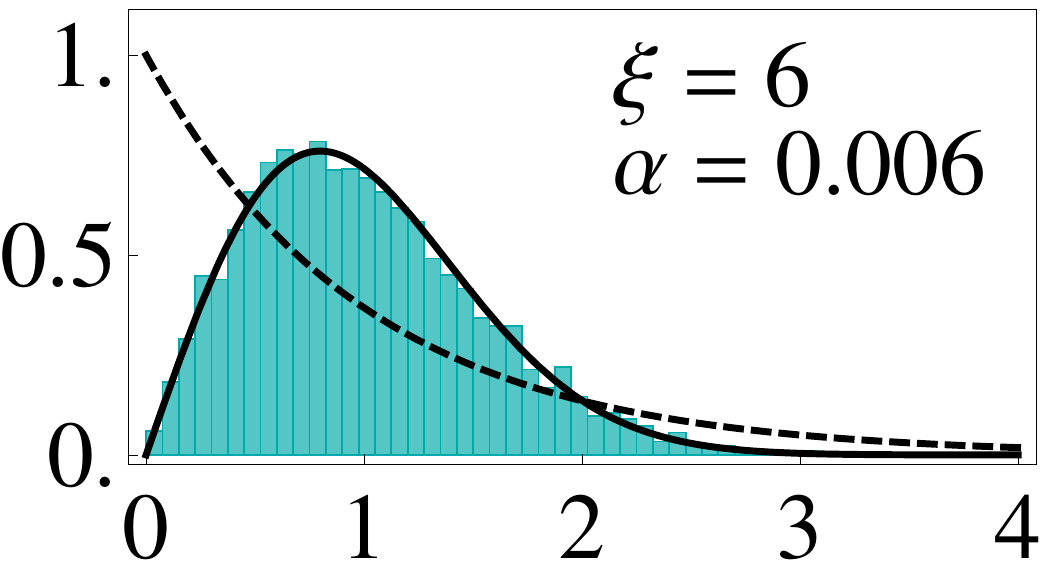}
\end{minipage}
\hfill
\begin{minipage}[t]{0.31\linewidth}
\centering
  \includegraphics[width=\linewidth]{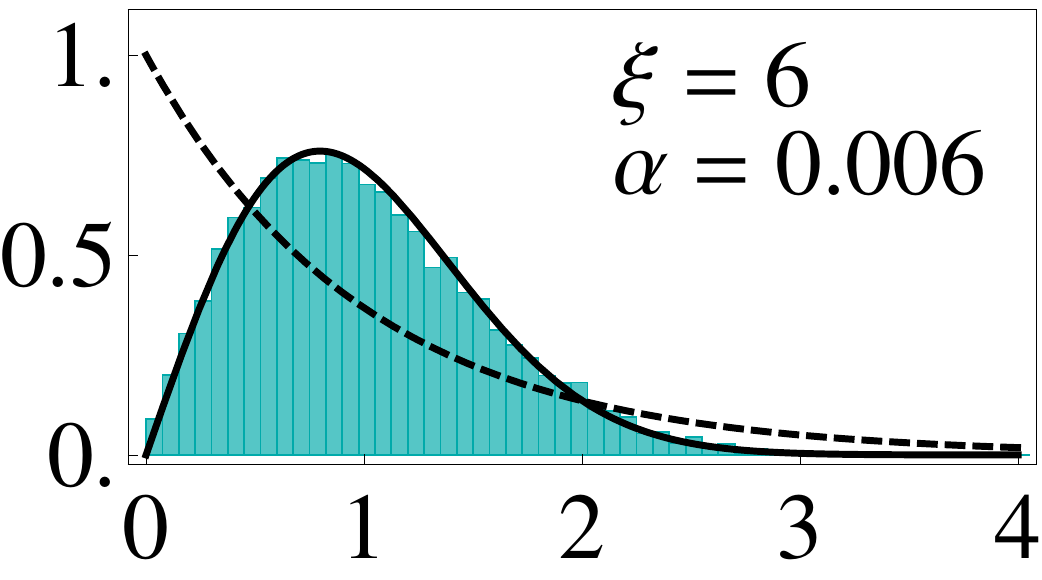}
\end{minipage}
\begin{minipage}[t]{0.35\linewidth}
\centering
  \includegraphics[width=\linewidth]{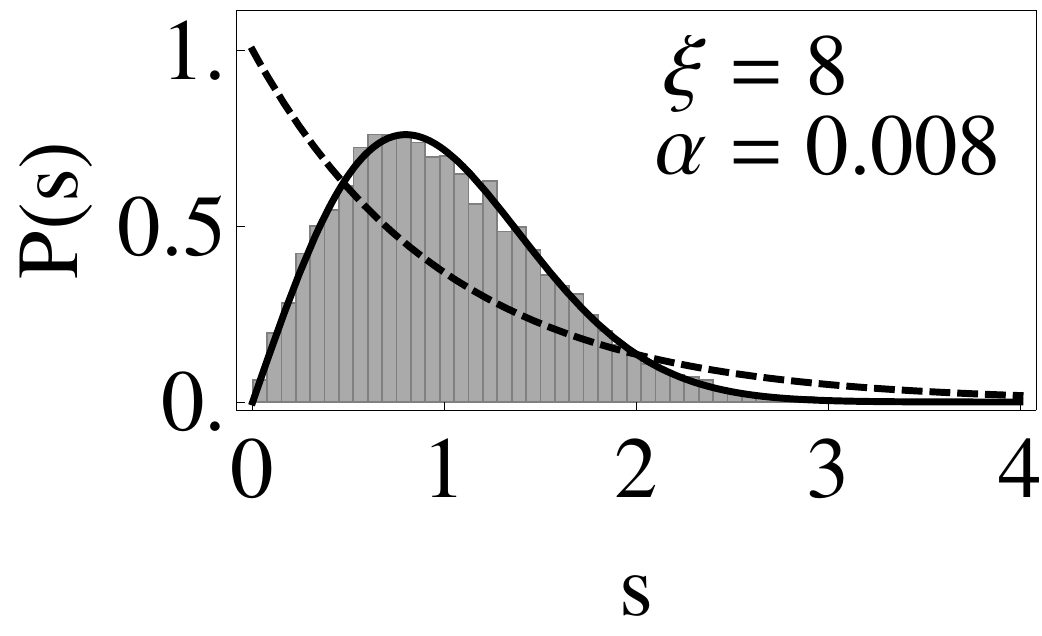}
\end{minipage}
\hfill
\begin{minipage}[t]{0.31\linewidth}
\centering
  \includegraphics[width=\linewidth]{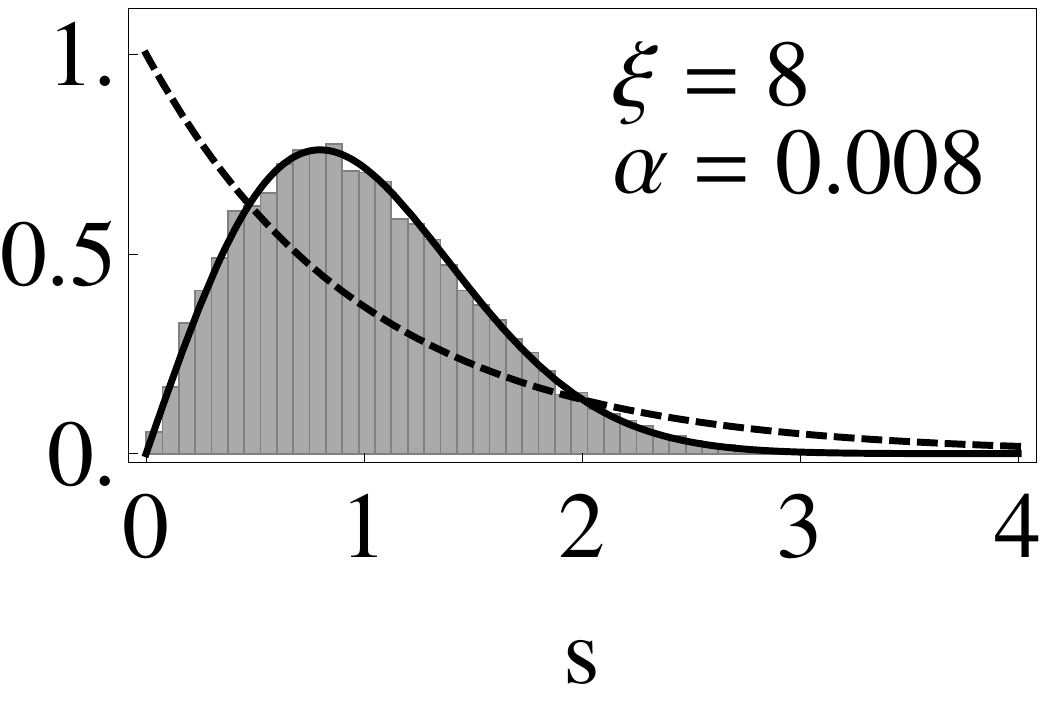}
\end{minipage}
\hfill
\begin{minipage}[t]{0.31\linewidth}
\centering
  \includegraphics[width=\linewidth]{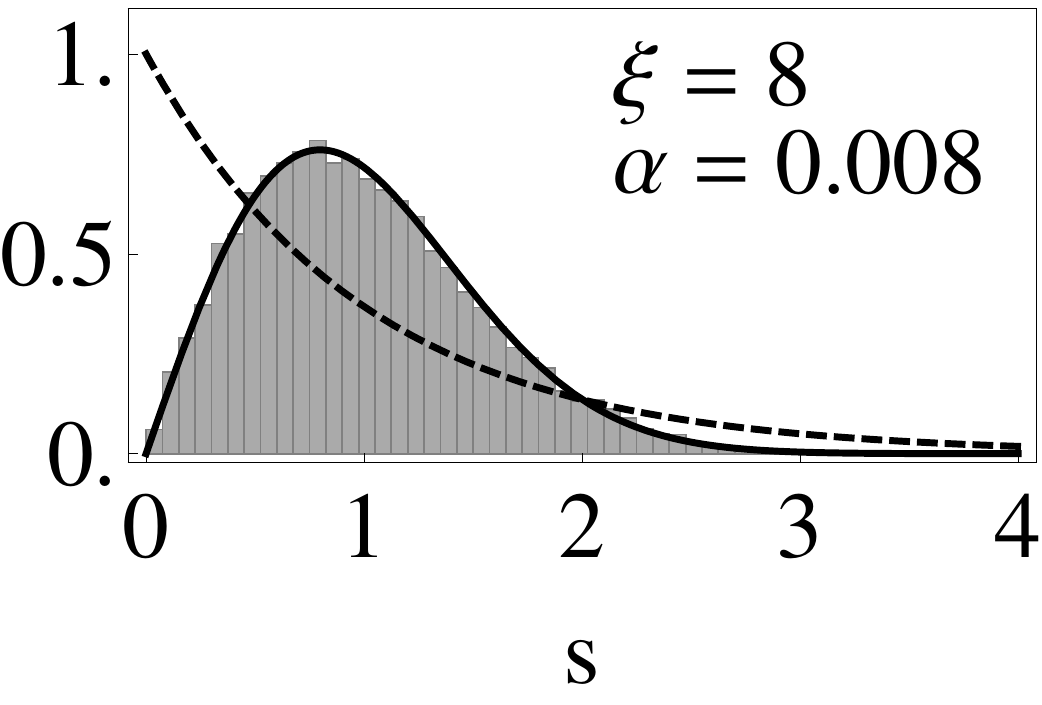}
\end{minipage}
\caption{Nearest neighbor spacing distribution (NNSD) for the same ensembles as in Fig.~\ref{comboPr}, for the three ER networks studied here: (a) fully random-weighted ER networks, (b) ER networks with random-weighted self-edges, and (c) standard ER networks. As in the $P(\tilde{r})$ distributions presented in Fig.~\ref{comboPr}, a very similar transition from Poisson to GOE statistics is observed as the average degree $\xi$ increases. Dashed and full lines correspond to the Poisson and GOE limits, respectively.}
\label{comboNNSD}
\end{figure}

Finally, we consider {\it standard ER networks}. In the standard ER random network model \cite{SR51,ER59,ER60}, the adjacency matrices are
random matrices with zeros in the main diagonal, and ones as non-vanishing off-diagonal elements. In other words, in the corresponding adjacency matrix, vertices and edges are represented with zeros and ones, respectively. It has been shown that the $P(s)$ of standard ER random networks is close to the Wigner-Dyson shape for large connectivity ($\alpha\to 1$)~\cite{BJ07,JB08,JB07,net01}. However, notice that $P(s)$ can not show the full Poisson to Wigner-Dyson transition since in the limit of vanishing connectivity ($\alpha=0$) the corresponding adjacency matrices are the null matrix.
\begin{figure}
\begin{center}
\begin{minipage}{0.93\linewidth}
\centering
\includegraphics[width=\linewidth]{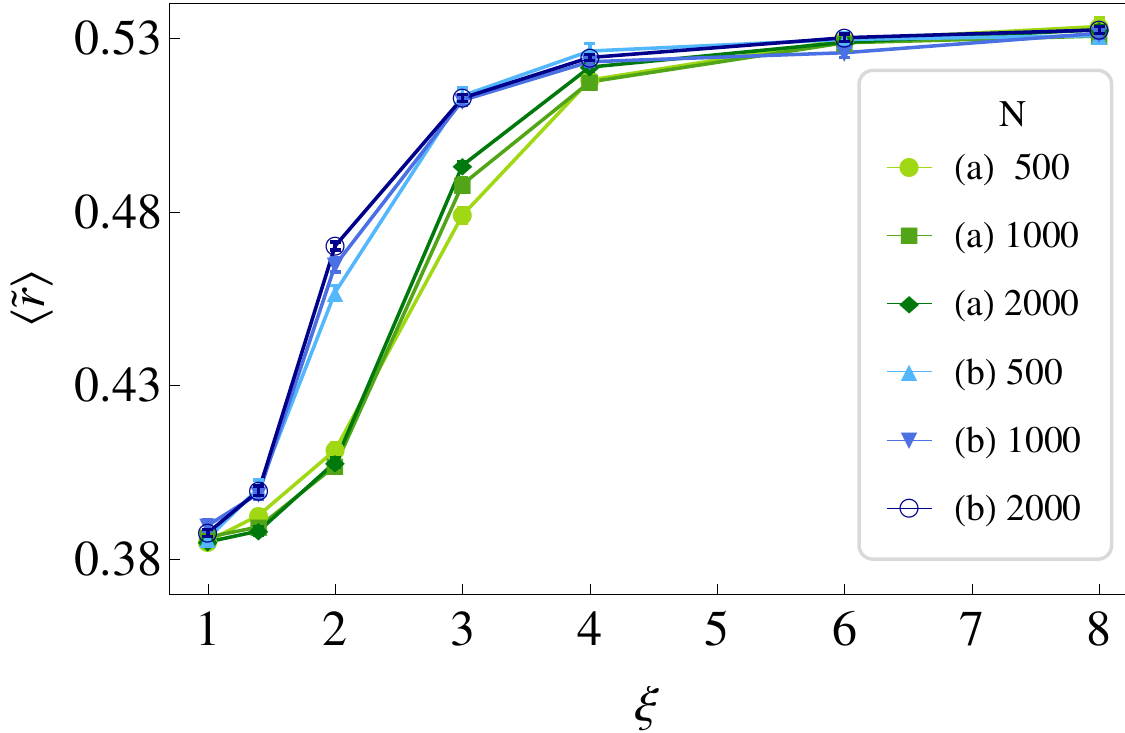}
\end{minipage}
\end{center}
\caption{Average of $\tilde{r}_n$ (Eq.~\ref{B3}), $\langle\tilde{r}\rangle$, for (a) fully random-weighted ER
networks (green), and (b) ER networks with random-weighted self-edges (blue), when $\xi=1,1.4,2,3,4,6,8$. A transition from Poisson to GOE statistics, which does not depend on $N$, is achieved as $\xi$, or equivalently the average network connectivity $\alpha$, increases. Error bars are not shown since they are much smaller than the symbol size.}
\label{rsN}
\end{figure}
\begin{figure}
\begin{center}
\begin{minipage}{0.8\linewidth}
\centering
\includegraphics[width=\linewidth]{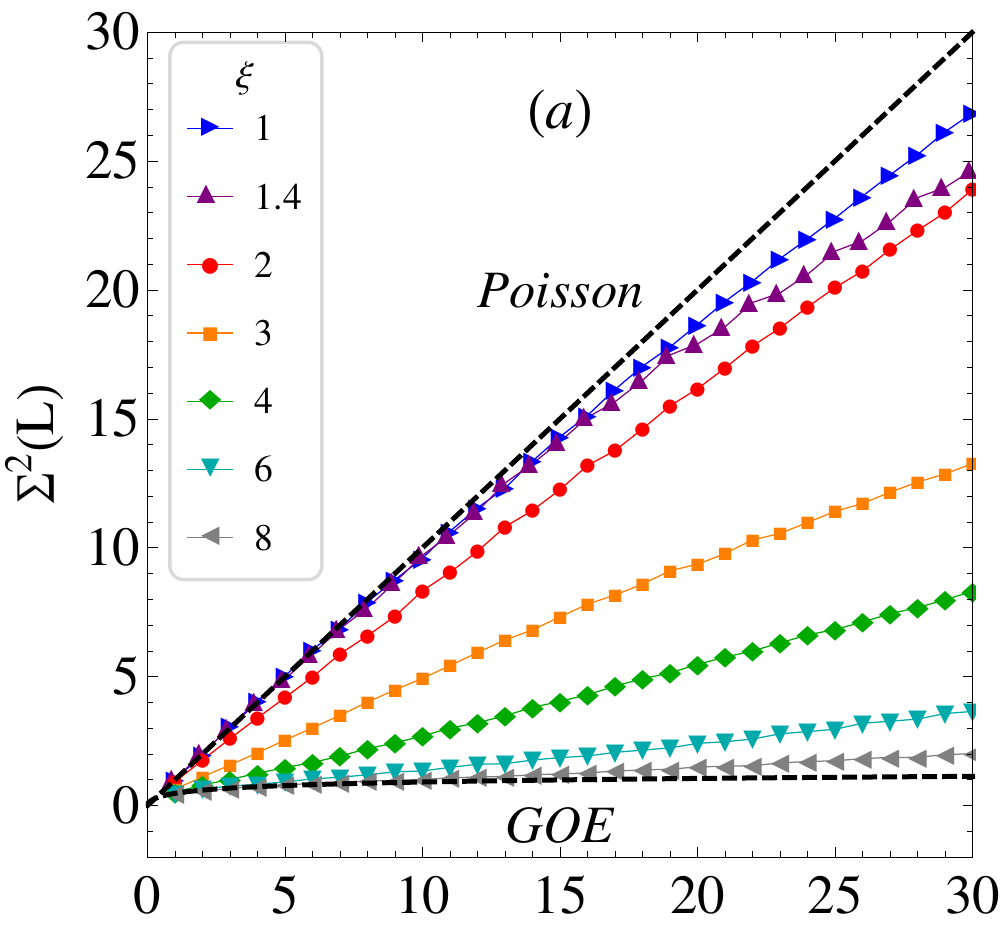}
\end{minipage}
\begin{minipage}{0.8\linewidth}
\centering
\includegraphics[width=\linewidth]{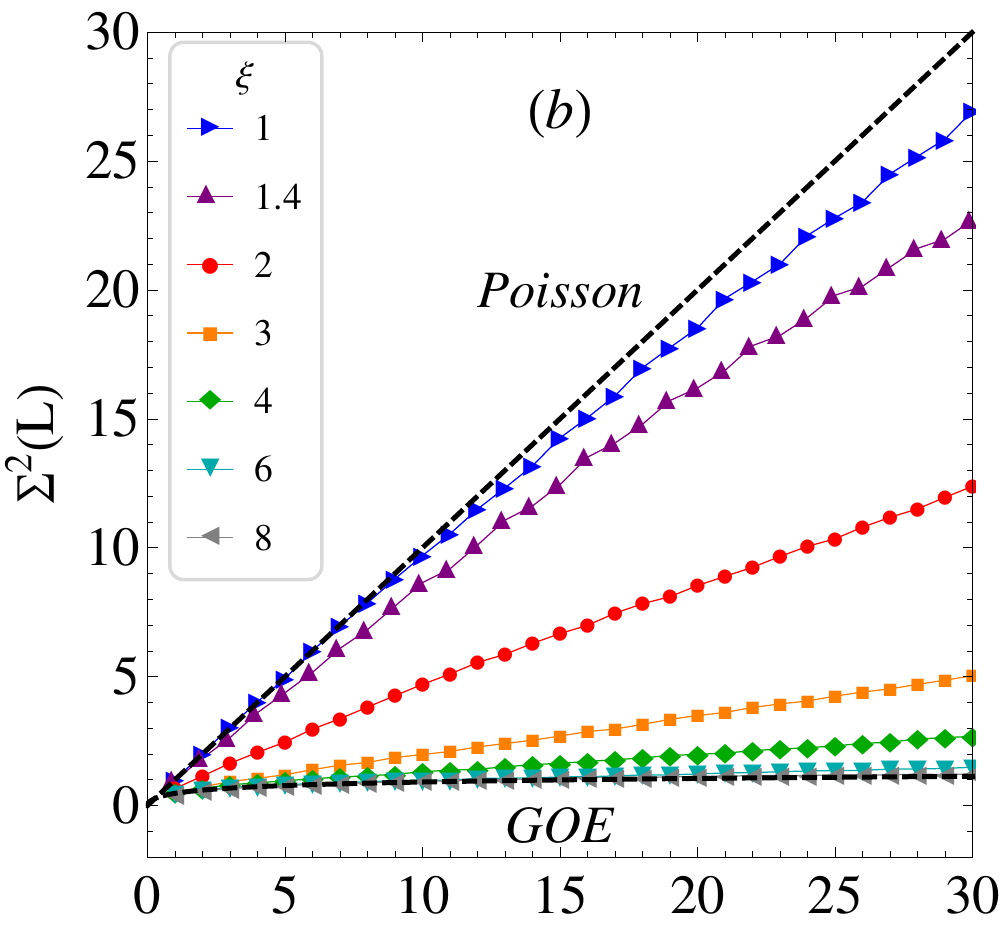}
\end{minipage}
\begin{minipage}{0.905\linewidth}
\raggedright
\includegraphics[width=0.925\linewidth]{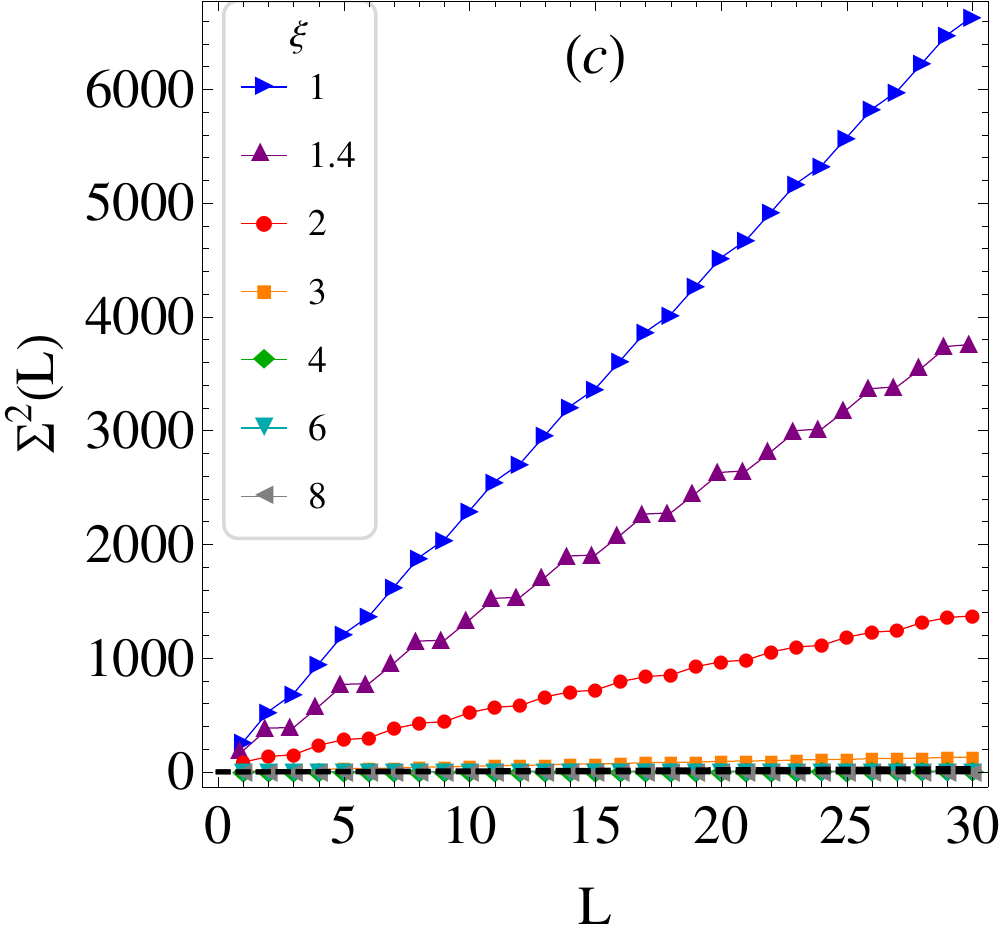}
\end{minipage}
\end{center}
\caption{Number variance, $\Sigma^2$, obtained for ensembles with $M=250$ eigenspectra of dimension $N=1000$, after performing a data-adaptive unfolding, corresponding to (a) fully random-weighted ER networks, (b) ER networks with random-weighted self-edges, and (c) standard ER networks. For fully random-weighted ER networks and ER networks with random-weighted self-edges we can observe that, as the values of $\xi$ increase, a transition from Poisson to GOE limits is achieved. The results correspond to ensemble averages.}
\label{comboS2}
\end{figure}
\begin{figure}
\begin{center}
\begin{minipage}{0.82\linewidth}
\centering
\includegraphics[width=\linewidth]{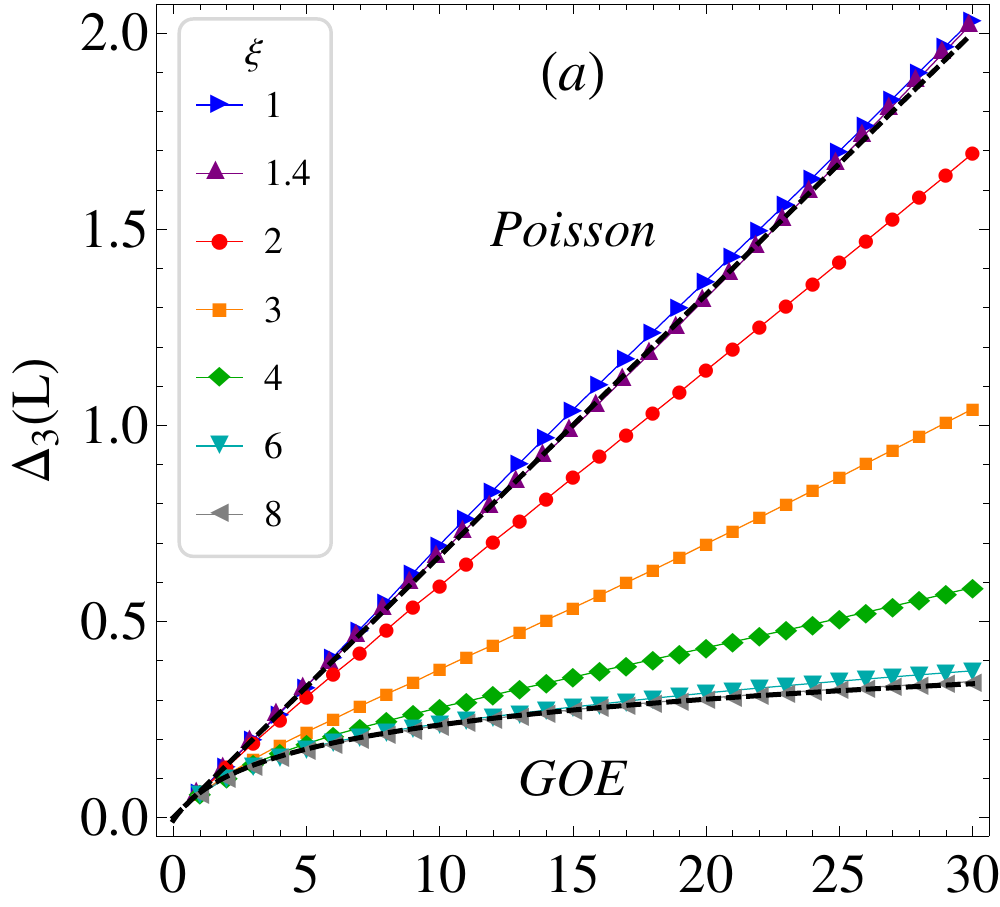}
\end{minipage}
\begin{minipage}{0.82\linewidth}
\centering
\includegraphics[width=\linewidth]{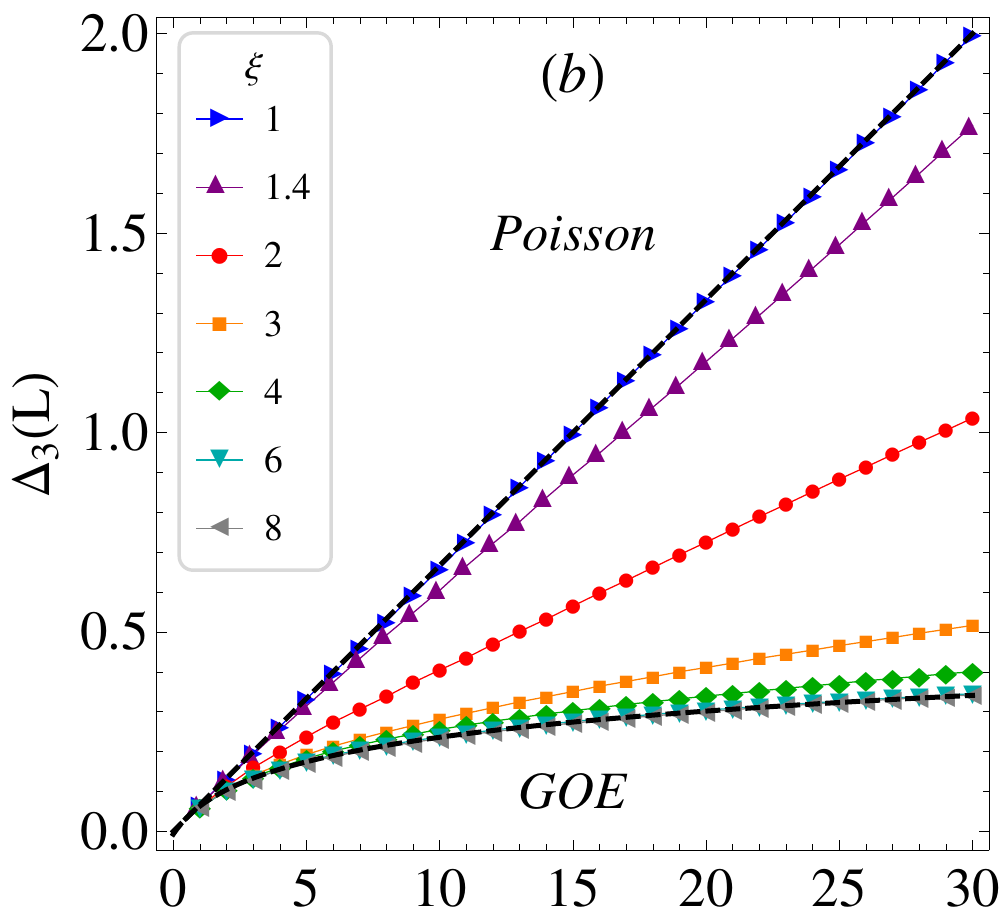}
\end{minipage}
\begin{minipage}{0.85\linewidth}
\raggedright
\includegraphics[width=0.985\linewidth]{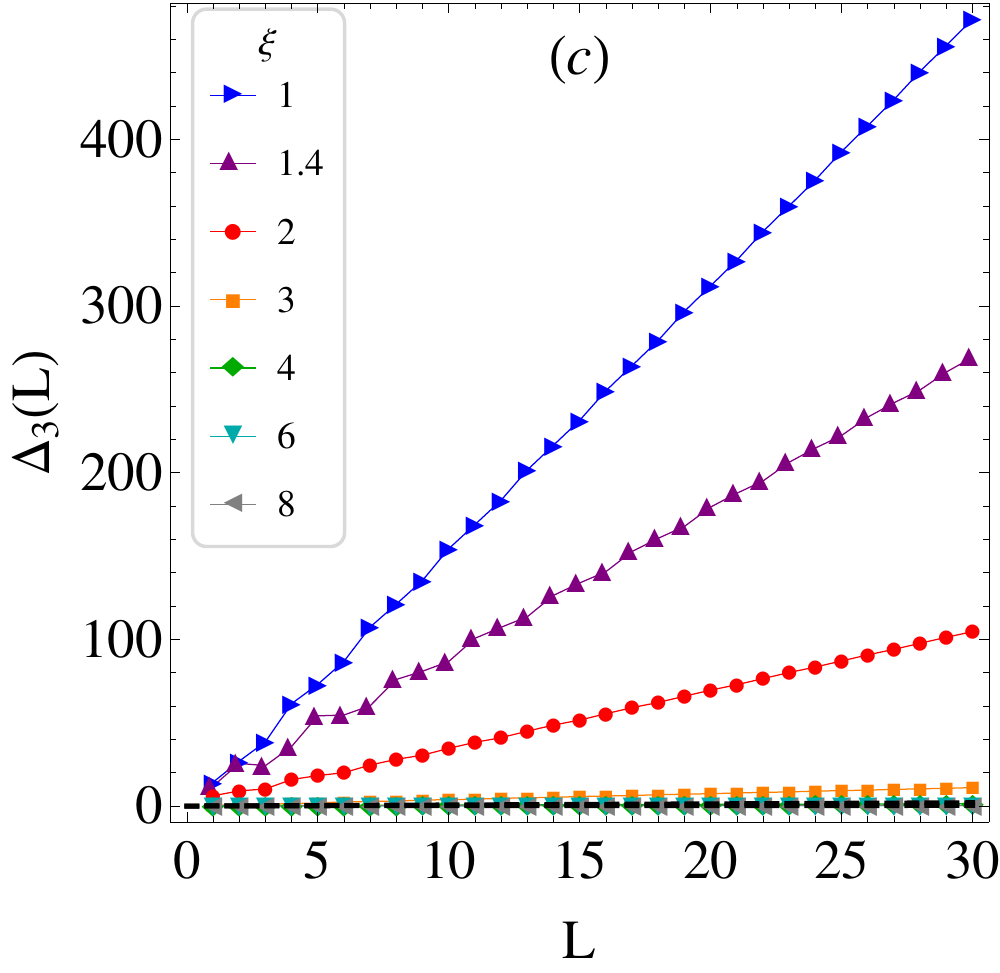}
\end{minipage}
\end{center}
\caption{$\Delta_3$ statistics obtained for the same ensembles as in Fig.~\ref{comboS2}, corresponding to (a) fully random-weighted ER networks, (b) ER networks with random-weighted self-edges, and (c) standard ER networks. Here, as for the number variance $\Sigma^2$, we can also see that for fully random-weighted ER networks and ER networks with random-weighted self-edges a transition from soft to rigid behavior is achieved as the values of $\xi$ increase. The results correspond to ensemble averages.}
\label{comboD3}
\end{figure}

Figure~\ref{comboSD} shows results for SVD applied to ensembles of $M = 125$, $250$, and $500$ eigenspectra $X^{(m)}(n)$. Each eigenspectrum contains $N = 500$, $1000$, and $2000$ eigenvalues, respectively, for: (a) fully random-weighted ER networks, (b) ER networks with random-weighted self-edges, and (c) standard ER networks. Only the central part of the eigenspectra were taken into account ($2.5\%$ of the lower and upper eigenvalues were discarded). As we can see in all the scree diagrams, $\lambda_1$ and $\lambda_2$ are orders of magnitude larger than the rest of partial variances, and they capture the major part of the total variance, $\lambda_{tot}=\sum_k\lambda_k$. As was explained in Sec.~\ref{SecI}, the corresponding eigenvectors $\vec{v}_k$, $k=1,2$ ($n_T=2$), constitute the basis states for the trend, $\overline{X}^{(m)}(n)$, of each ensemble defined by a given pair $\alpha$ and $N$. On the other hand, the (much smaller) higher-order partial variances $\lambda_k$, with $k=3,\ldots,r$, behave as a power law, except for (a) $\xi=6$ and 8, (b) $\xi\ge 3$, and (c) $\xi\ge 2$. In these cases, the scale invariance of the fluctuations is broken and a crossover in the scree diagrams, between the Poisson and GOE limits, appears. For a given $N$, the location of the crossover $k_x$ shifts towards higher order-numbers as $\alpha$ decreases, in correspondence with Ref.~\cite{jac01}, where also  $k_x\propto\sqrt{N}$ was found. In the following, results will be presented for intermediate ensemble sizes: $M=250$ eigenspectra and $N=1000$ eigenvalues.

Figure~\ref{comboPr} shows the distribution $P(\tilde{r})$ obtained for the three ER random network models considered here. As we can see, for fully random-weighted ER networks and ER networks with random-weighted self-edges, a transition from Poisson to GOE statistics is observed as the values of $\xi$ increase. The transition for standard ER networks is not smooth since this model does not reproduce the Poisson limit when $\alpha\to 0$. Indeed, the large peak of $P(\tilde{r})$ at $\tilde{r}=0$ is a signature of spectral degeneracies appearing when $\alpha\to 0$. In Fig.~\ref{rsN}, we present the average values of $\tilde{r}_n$ (Eq.~\ref{B3}), $\langle\tilde{r}\rangle$, for: (a) fully random-weighted ER networks, and (b) ER networks with random-weighted self-edges, when $\xi=1,1.4,2,3,4,6,8$. In this way, we can see that the transition from Poisson to GOE statistics, does not depend on $N$. Moreover, we can also appreciate that this transition occurs faster for the ER networks with random-weighted self-edges than for fully random-weighted ER networks. If we compare results from Fig.~\ref{comboSD} with those of Fig.~\ref{comboPr}, we can observe that although the $P(\tilde{r})$ distribution shows already a GOE shape for $\xi=4$, the corresponding result obtained for the scree diagrams suggests still Poisson behavior. This illustrates the importance to study not only short-range correlation statistics but to include also large-range fluctuation measures.

Now, we compare the previous results with those obtained for the traditional spectral fluctuation measures used in RMT. Fig.~\ref{comboPS} shows the power spectra, $P(f)$, of the $\delta_n$ statistics, after performing a data-adaptive unfolding for: (a) fully random-weighted ER networks, (b) ER networks with random-weighted self-edges, and (c) standard ER networks. As we can see, in the case of fully random-weighted ER networks and ER networks with random-weighted self-edges, the power spectra also exhibit a crossover between the Poisson and GOE limits. However, although such crossover is very similar to that observed in the fluctuation modes of Fig.~\ref{comboSD}, it can be better appreciated in the scree diagrams. For standard ER networks, because of the nature of the eigenspectra, the $\delta_n$ statistics can not be properly calculated, and therefore the corresponding $P(f)$ obtained does not exhibit a clear crossover. This is another advantage of the scree diagram over the $P(f)$.

In Fig.~\ref{comboNNSD} we present the results of NNSD for: (a) fully random-weighted ER networks, (b) ER networks with random-weighted self-edges, and (c) standard ER networks. As we can see, these results characterize the transition between the Poisson and GOE limits, in a very similar way to those obtained for the $P(\tilde{r})$ distribution, which does not need a previous unfolding, see Fig.~\ref{comboPr}. In Fig.~\ref{comboS2} and Fig.~\ref{comboD3}, we show the corresponding results for $\Sigma^2$ and $\Delta_3$ statistics, respectively. For fully random-weighted ER networks and ER networks with random-weighted self-edges, a good characterization of the transition from soft (Poisson) to rigid (GOE) behavior, as $\xi$ increases, is achieved. Like for $P(\tilde{r})$ and $P(s)$ distributions, we can appreciate that $\Sigma^2$ and $\Delta_3$ statistics does not converge to the Poisson limit, as expected for the standard ER networks when $\alpha\to 0$, since the corresponding adjacency matrices are the null matrix.

Although NNSD is a short-range spectral fluctuation measure, and $\Sigma^2$ and $\Delta_3$ are long-range spectral fluctuation measures, both types of measures indicate approximately the same statistics for the system under study. For example, in the case of fully random-weighted ER networks, when $\xi=1$ and 1.4, short and long-range fluctuation measures indicate Poisson statistics, while for $\xi=6$ and 8, the results indicate GOE behavior. For the rest values of $\xi$, we can appreciate intermediate statistics. For ER networks with random-weighted self-edges we have that when $\xi=1$ we still found Poisson behavior, and when $\xi\ge 4$ the system is in the GOE limit. In this sense, the scree diagram offers additional information to that provided by the standard spectral fluctuation measures of RMT. Moreover, we should remember that the calculation of the NNSD, $\Sigma^2$, $\Delta_3$ and $\delta_n$ statistics requires carrying out an unfolding procedure, which, if is not properly performed, could lead to the introduction of artifacts.

\section{Conclusions}
\label{Conclusions}

We have interpreted network spectra as time series. In particular, we studied three different representations of ER random networks: fully random-weighted ER networks, ER networks with random-weighted self-edges, and standard ER networks. In all cases, SVD was applied to decompose the corresponding spectra in trend and fluctuation normal modes. By using the trend modes, we performed a data-adaptive unfolding of the network spectra in order to calculate the NNSD, $\Sigma^2$, $\Delta_3$ and $\delta_n$ statistics from RMT. In this way, we contrasted the results obtained without implementing any unfolding procedure (SVD), with those obtained after unfolding (RMT). 

We characterized the long-range spectral correlations by means of the fluctuation modes, and through the power spectrum of the $\delta_n$ statistics. Both approaches showed a crossover, not identified before, from the Poisson and the GOE statistics, as the average degree of ER networks increases. In all cases, it was possible to calculate the scree diagram, unlike the power spectrum which could not be properly calculated for standard ER networks, due to the nature of the spectra. Moreover, we found that SVD is able to detect fine details in the crossover that standard RMT measures can not. We also observed a transition between Poisson and GOE behavior in the short-range spectral correlations, which were measured by the $P(\tilde{r})$ and $P(s)$ distributions. We found that such transition is independent of the network size, and it occurs faster for the ER networks with random-weighted self-edges, than for fully random-weighted ER networks.

It is important to emphasize that SVD is the first method to study long-range fluctuation statistics without unfolding, and after the detailed comparison of short and long-range correlation RMT measures made here, we conclude that SVD is not only a reliable tool to characterize random network spectra, but it offers additional information to that provided by the standard spectral fluctuation measures of RMT. This can be particularly important when studying the spectra of binary adjacency matrices, that are not well handled by traditional RMT measures.

\section*{Acknowledgements}

Financial funding for this work was supplied by the Direcci\'on General de Asuntos del Personal Acad\'emico (DGAPA) from the Universidad Nacional Aut\'onoma de Mexico (UNAM) with grants PAPIIT IA105017 and IA102619. We are thankful for the Newton Advanced Fellowship awarded to RF by the Academy of Medical Sciences through the UK Governments Newton Fund program.

\end{document}